\documentclass{article}

\usepackage{arxiv}

\usepackage[utf8]{inputenc} 
\usepackage[T1]{fontenc}    
\usepackage{hyperref}       
\usepackage{url}            
\usepackage{booktabs}       
\usepackage{amsfonts}       
\usepackage{nicefrac}       
\usepackage{microtype}      
\usepackage{lipsum}
\usepackage{graphicx}
\usepackage[table]{xcolor}

\usepackage[numbers,square]{natbib}
\usepackage{tikz}
\usepackage{booktabs}
\usepackage{multirow}
\usepackage{makecell}
\usepackage{capt-of}
\usepackage{wrapfig}
\newcommand{\scorepair}[2]{#1 \hspace{0.5em} #2}
\newcommand{\scoretriple}[3]{#1 \hspace{0.25em} #2 \hspace{0.25em} #3}
\usepackage{subcaption}

\usepackage{
lmodern,
pgfplots,
tikz,
}
\usepgfplotslibrary{groupplots}

\pgfplotsset{
   compat=1.16,
   legend entry/.initial=,
}

\definecolor{bblue}{HTML}{4F81BD}
\definecolor{rred}{HTML}{FFB303}
\definecolor{ggreen}{HTML}{9BBB59}
\definecolor{igreen}{HTML}{579c35}
\definecolor{ppurple}{HTML}{9F4C7C}

\usepackage{bigfoot}

\DeclareNewFootnote{AAffil}[arabic]
\DeclareNewFootnote{ANote}[fnsymbol]

\usepackage{etoolbox}
\makeatletter
\patchcmd\maketitle{\def\@makefnmark{\rlap{\@textsuperscript{\normalfont\@thefnmark}}}}{}{}{}
\makeatother

\makeatletter
\def\thanksAAffil#1{
  \footnotemarkAAffil\protected@xdef\@thanks{\@thanks%
        \protect\footnotetextAAffil[\the \c@footnoteAAffil]{#1}}%
}
\def\thanksANote#1{%
  \footnotemarkANote%
  \protected@xdef\@thanks{\@thanks%
        \protect\footnotetextANote[\the \c@footnoteANote]{#1}}%
}
\makeatother

\author{
 Viktor Schlegel%
 \thanksAAffil{Imperial College London, Imperial Global Singapore}$\ ^{,}$\hspace{-0.15cm}
 \thanksAAffil{University of Manchester, Department of Computer Science}$\ ^{,}$\thanksANote{Corresponding author: \texttt{v.schlegel@imperial.ac.uk}}
   \And 
 Anil A Bharath%
 \footnotemarkAAffil[1]$\ ^{,}$\hspace{-0.15cm}
 \thanksAAffil{Imperial College London, Department of Bioengineering}
  \And
 Zilong Zhao%
  \thanksAAffil{betterdata.ai, Singapore}\hspace{0.02cm}$\ ^{,}$\thanksAAffil{National University of Singapore, Asian Institute of Digital Finance}
  \And
 Kevin Yee%
  \footnotemarkAAffil[4]
}

\begin{document}
\let\WriteBookmarks\relax
\def\floatpagepagefraction{1}
\def\textpagefraction{.001}

\title{Generating Synthetic Data with Formal Privacy Guarantees: State of the Art and the Road Ahead}

\maketitle
\begin{abstract}

Privacy-preserving synthetic data offers a promising solution to harness segregated data in high-stakes domains where information is compartmentalized for regulatory, privacy, or institutional reasons. This survey provides a comprehensive framework for understanding the landscape of privacy-preserving synthetic data, presenting the theoretical foundations of generative models and differential privacy followed by a review of state-of-the-art methods across tabular data, images, and text. Our synthesis of evaluation approaches highlights the fundamental trade-off between utility for down-stream tasks and privacy guarantees, while identifying critical research gaps: the lack of realistic benchmarks representing specialized domains and insufficient empirical evaluations required to contextualise formal guarantees.

Through empirical analysis of four leading methods on five real-world datasets from specialized domains, we demonstrate significant performance degradation under realistic privacy constraints ($\epsilon \leq 4$), revealing a substantial gap between results reported on general domain benchmarks and performance on domain-specific data. 
These challenges underscore the need for robust evaluation frameworks, standardized benchmarks for specialized domains, and improved techniques to address the unique requirements of privacy-sensitive fields such that this technology can deliver on its considerable potential.

\end{abstract}












\section{Introduction} 
\subsection{Motivation} 
High-stakes domains, such as healthcare and financial services, face mounting challenges in leveraging data effectively to support expert decision-making. 
While recent advances in artificial intelligence and data-driven decision support have enabled data-driven decision making~\cite{Sutton2020AnSuccess}, the arising solutions are tailored to specific application scenarios, and require significant effort and domain-specific knowledge. 
Conversely, Generative AI demonstrates promising capabilities in making sense of complex and unstructured data and generalise to new scenarios with little context, emerging as a general purpose technology~\cite{Eloundou2023GPTsModels}\footnote{Though this notion is contested~\cite{Rogers2023Position:Footnotes}.}. 
While this is true in in the public domain, the application of AI systems in domains where data remains highly compartmentalized, has been shown to face unique challenges. 

The performance and reliability of data-driven algorithms often deteriorate in those domains, exemplifying the ``jagged frontier'' of AI~\cite{DellAcqua2023NavigatingQuality}, where counter-intuitive behaviour (e.g., LLMs perform well at arithmetic calculations \cite{Cobbe2021TrainingProblems} while failing at basic counting tasks~\cite{Ball2024CanCapabilities,Fu2024WhyLetters}) makes it difficult to develop reliable mental models of system capabilities. This leads domain experts to hesitate in adopting these technologies for critical workflows despite their potential benefits~\cite{DellAcqua2023NavigatingQuality}.
    
The strong progress in AI technology can be attributed in large part to standardized benchmarks and publicly available resources, which have fostered healthy competition and collaborative improvement within the research community~\cite{Deng2009ImageNet:Database,Wang2019,Chiang2024ChatbotPreference}. 
These shared resources and metrics have been instrumental in understanding the limitations of existing technology and making targeted improvements for successive iterations (for instance, fine-tuning language models on a collection of publicly available resources for various tasks helps language models generalise to many unseen tasks that can be expressed in language~\cite{Chung2024ScalingModels-abbr}), and, together with the availability of computational resources and web-scale amount of training data, resulted in the AI breakthroughs we are witnessing today~\cite{OpenAI2023GPT-4Reportb-abbr}.

Therefore, for experts in high-stakes domains to benefit from these breakthroughs in similar fashion, there is a need to represent their tasks, including the corresponding data, in publicly available resources that capture the complexity and nuance of sensitive domains like healthcare and finance. 
However, curating such datasets poses unique challenges, as the underlying data typically cannot be shared due to privacy concerns, regulatory requirements, and institutional boundaries. 

Popular mechanisms for controlling disclosure when sharing data exist, such as anonymisation via redaction~\cite{Lison2021AnonymisationDirections}, or $k$-anonymity~\cite{sweeney2002k}, where a level of anonymity is achieved by binning individuals with similar characteristics and only report those bins rather than the characteristics themselves (e.g., the age of 78 might be reported as belonging to the 70-80 bin). However, they have been shown to be vulnerable to re-identification attacks~\cite{Machanavajjhala2007L-diversity:K-anonymity} and either provide no formal guarantees or are too specific to modalities (e.g., it is hard to imagine that binning text with $k$-anonymity approaches would keep the result interpretable to humans~\cite{Maeda2016FastK-anonyminity}). Another approach is to conduct statistical disclosure analyses~\cite{Hundepool2012StatisticalControl}, which does not provide formal guarantees regarding \emph{the sharing mechanism} itself, rather than the released data (and, again, has limited transferability to unstructured data).

Using generated synthetic facsimile in lieu of the sensitive data emerges as a promising solution to this challenge, as state-of-the-art generative AI methods show increased promise to generate realistic data~\cite{Schick2021GeneratingModels,Rombach2022High-resolutionModels,ctabgan}. 
However, for synthetic data to be useful, it must faithfully represent the statistical patterns and relationships present in those compartmentalised real-world datasets.
Without real-world representativeness, performance of models trained on (purely) synthetic data will deteriorate~\cite{Shumailov2024AIData} and fail to generalise to real-world applications and analyses based on synthetic data could lead to erroneous conclusions. 
This requirement for representativeness is seemingly at odds with the strict privacy protection laws that prevent the sharing of sensitive data. 
Differential privacy~\cite{dwork2014algorithmic} offers a potential resolution to this tension by providing formal guarantees about the maximum influence any individual entry in a private dataset can have on the resulting synthetic data, thus exacerbating reidentification efforts. 
This approach enables organizations to generate high-quality representative synthetic data while maintaining provable privacy guarantees, benefitting from the research advancements of the wider community in adapting general purpose technology to their application domains.

The rapid advancement of techniques for privacy-preserving synthetic data generation has led to a steadily growing body of research, spanning theoretical frameworks \cite{dwork2014algorithmic,Dong2022GaussianPrivacy,Mironov2017RenyiPrivacy}, algorithmic innovations \cite{Abadi2016DeepPrivacy,bu2023differentially,Yu2021DifferentiallyModelsb}, and practical applications\footnote{e.g. \url{https://desfontain.es/blog/real-world-differential-privacy.html} for various DP data releases} across diverse domains. This accelerated activity calls for a comprehensive survey to charter the emerging landscape of this novel research field. We heed this call in this paper, serving multiple purposes: \textbf{first}, we aim to provide researchers with a structured entry point into this emerging field, helping them navigate the landscape of existing methods, understand key theoretical foundations, and identify promising directions for future work; the survey focuses on  the recent advancements of generative AI, which promise the synthesis of rich unstructured datasets. \textbf{Second}, we seek to make data holders across industries aware of the latest technological capabilities, enabling them to make informed decisions about implementing synthetic data solutions while maintaining privacy guarantees. \textbf{Finally}, with this survey, we intend to initiate and inform dialogue among policymakers regarding the regulation of synthetic data, helping to establish frameworks that balance innovation with privacy protection in an era where data sharing is increasingly critical for technological progress.

\begin{figure}[h!]
\centering
\includegraphics[width=0.95\linewidth]{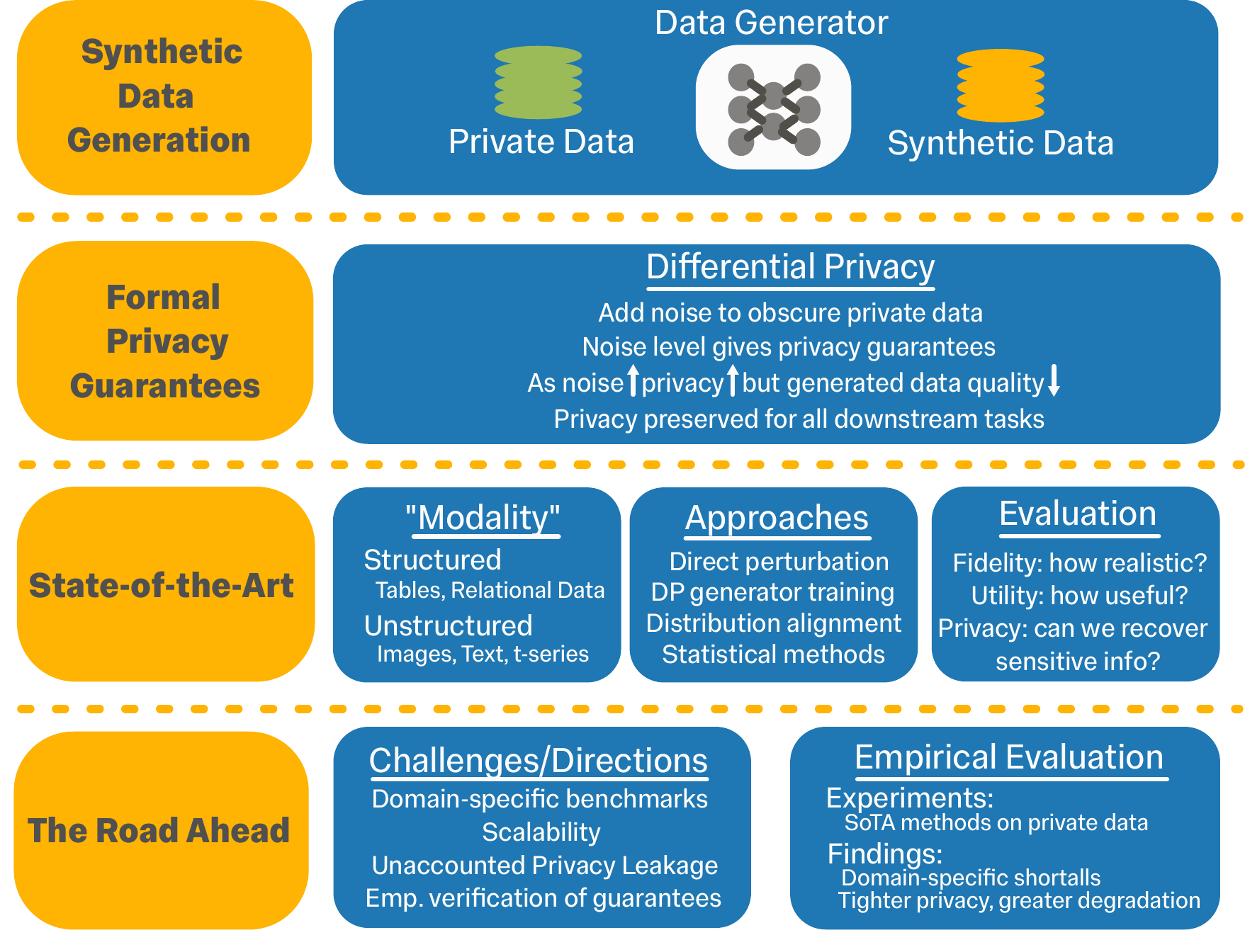}
\label{fig:surveyview}
\caption{This paper considers the techniques available for synthetic data generation according to the principles of generative modelling, focussing on the need to preserve privacy, so that sensitive content of private source information does not leak into synthetic data, or down-stream tasks based on that data. The formal privacy guarantees of Differential Privacy are attractive. We evaluate techniques for different modalities, and approaches, and evaluate the outputs. Based on a series of experiments, future directions for improvement are suggested.}
\end{figure}

\subsection{Scope, Methodology and Related Surveys}
In this survey, we focus on dataset-level privacy where the risk of identifying any individual that contributed to that dataset can be quantified. Notably, we do not include redaction~\cite{Lison2021AnonymisationDirections}, obfuscation~\cite{Li2021DifferentiallyManipulation} or ``rewriting''~\cite{Krishna2021ADePT:Transformation} methods that preserve the link between original and synthetic data points. To protect the privacy of individuals from re-identification attacks on data manipulated this way would require the result of the obfuscation mechanism  to be indistinguishable from any kind of input of the same modality. \citet{Habernal2021WhenDetail} gives the following example: Observing the result \emph{``on april first i need a ticket from tacoma to san jose departing before 7am''} of the obfuscation process should be roughly equiprobable to either obfuscating \emph{``on april first i need a flight going from phoenix to san diego''} or \emph{``monday morning i would like to fly from columbus to indianapolis''}, or indeed, any other such utterance. Because of this and other challenges~\cite{Bassolas2019HierarchicalLivability}, researchers relax (sometimes wrongly~\cite{Habernal2021WhenDetail}) the assumptions (e.g., \mbox{\citet{Li2021DifferentiallyManipulation}} preface their claim about reidentification prevention with a condition: ``if you do not already know who they are''). Instead, we focus on approaches that create synthetic data from a ``trusted curator'', i.e. an entity (e.g. a hospital, a bank, etc) that has access to a set of records and want to protect the privacy of each individual contributing to that dataset. Here the focus is on privacy preservation for data release.

In terms of methodology, rather than performing an exhaustive and structured survey, we first provide the reader with the necessary background (Section~\ref{sec:background}). We then map out the field of synthetic data generation with privacy guarantees and highlight key contributions, pointing to related work and surveys (such as, e.g.,~\citet{Gadotti2024Anonymization:Privacy}, for a broader view on anonymisation techniques for data sharing) where appropriate in Section~\ref{sec:methods}. We pay specific attention to the evaluation methods for synthetic data generation in Section~\ref{sec:evaluation} and conclude with set of challenges and future directions in Section~\ref{sec:challenges}. We validate some of the challenges empirically in Section~\ref{sec:validation} and conclude with Section~\ref{sec:conclusion}.

\section{Background} 
\label{sec:background}
\subsection{Simple Privacy Measures}
The idea of modifying data in order to remove traits that reveal the identify of distinct individuals can take many forms. It is useful to conceptualise the different approaches as existing somewhere in a space of complexity, severity of attribute obfuscation (including deletion of specific fields of record) and the degree of loss of usefulness of the data for some end purpose. 

An obvious demonstration is found in the well-known idea of obfuscating dates of birth by deletion of month or day of birth from records. This amounts to binning people's ages into year-wide bands. For medical data, privacy might be attained by binning age into decade wide bins, which still retains a key variable for patient stratification, patient age, albeit approximately.

Binning -- rigorously studied by Sweeney \cite{sweeney2002k} -- can still fail to preserve privacy. For instance, here may be few individuals in the records of, say, a small town, that are between the ages of 90 and 99. Should some records show the presence of an embarrassing condition, an engineer or data scientist tasked with estimating the prevalence of that medical condition within different age ranges might be able to readily identify the afflicted individual; this is beyond the (privacy respecting) remit of simply estimating prevalence. The violation of privacy arises because of the detailed nature of the task and the joint data distribution across variables -- i.e. the distribution of condition $X$ according to decade-binned age -- and the sparsity of data in some age ranges. It is telling that the US HIPAA (Health Insurance Portability and Accountability Act) of 1996 recommends ``capping'' ages at 90 as one of the requirements to combat identification of individuals in health records \cite{moffatt2024best}.

The example allows us to highlight a further feature of anonymisation by binning: if there are \textit{two} individuals in the 90-99 age range, and if the number of people that have the condition is 1, then there is a 50\% chance of the condition being held by each of the nonagenarians. This illustrates a key characteristic of privacy: that the revealing of protected information about certain characteristics is dependent on the distribution and number of individuals across other traits of interest. Using wider bins might be seen as protecting against this problem, but comes at the expense of a loss of usefulness, associated with a broadening of the binning conditions. Potential loss of usefulness or \textit{utility} can be characterised through measures of information loss (see for example \cite{domingo-ferrer:2009}).

An approach to protecting privacy that exploits stochasticity, developed in the social sciences to collect sensitive information from participants, is reminiscent of the Knights and Knaves puzzle \cite{cook:2006}, with a twist (or perhaps, a ``flip''); it may be used in conducting sensitive surveys. Participants in a study are asked to toss a coin. If the coin shows tails, they respond truthfully to a personal question asking them if they have undertaken or engaged in some embarrassing or illegal activity. If the coin shows heads, they flip the coin again and respond with a ``yes'' if heads, and a ``no'' if tails. In this approach, a single participant that answers ``yes'' has not necessarily engaged in anything illegal.  A researcher who wishes to estimate $x$, the proportion of a population that engages in an illegal or embarrassing activity, need only ask a sufficient number of participants, $N$, to take the survey with the randomisation mechanism. To calculate $x$, a quarter of responses is discounted as random ``yes'' responses and assuming that the proportion of true ``yes'' answerers is equal in both truthfully and randomly answering groups, the true value of $x$ should be double that of the observed number of ``yes'' responses after discounting the random ones. An estimate of $x$ can then be extracted from the number of  ``yes'' answers ($N_y$) after some arithmetic simplification:
$$
x = 2\frac{N_y}{N} - \frac{1}{2}
$$
It turns out that some form of randomness is one of the key mechanisms to create sources of  differentially private data. We return to this in Subsection~\ref{subsec:DPIntro}.

\subsection{Machine Learning and Embedding Spaces}
When the data consist of a small number of directly observed variables, the protection of sensitive data through binning, or adding elements of randomness to observations, clearly offers some mechanisms of privacy. How do we extend these ideas when the data are multivariate, and the mapping between observations and an inferred answer is vastly more complex? This situation is encountered when data-driven AI is used.

Recall that many problems addressed by deep learning -- including language processing and image-related tasks -- harness the idea of an embedding space. An embedding space is closely related to the idea of a latent space. The term latent space is particularly relevant when there are latent variables that compactly capture the distribution of input data. Embedding spaces are similar (in some cases, one might argue, identical), but are more commonly associated with higher-dimensional encoding of data that lie on a lower-dimensional manifold. For the sake of this discussion, we do not make a strong distinction, and instead consider training a network to perform some well-defined task, so that it explicitly or implicitly encodes a dataset of, say, patient characteristics and diagnoses into an embedding space; see for example \cite{chushig2022learning} for an example of embedding healthcare records based on denoising autoencoders.

\begin{figure}[ht!]
\centering
\includegraphics[width=0.45\linewidth, trim={5cm 12cm 2cm 3cm}, clip=true]{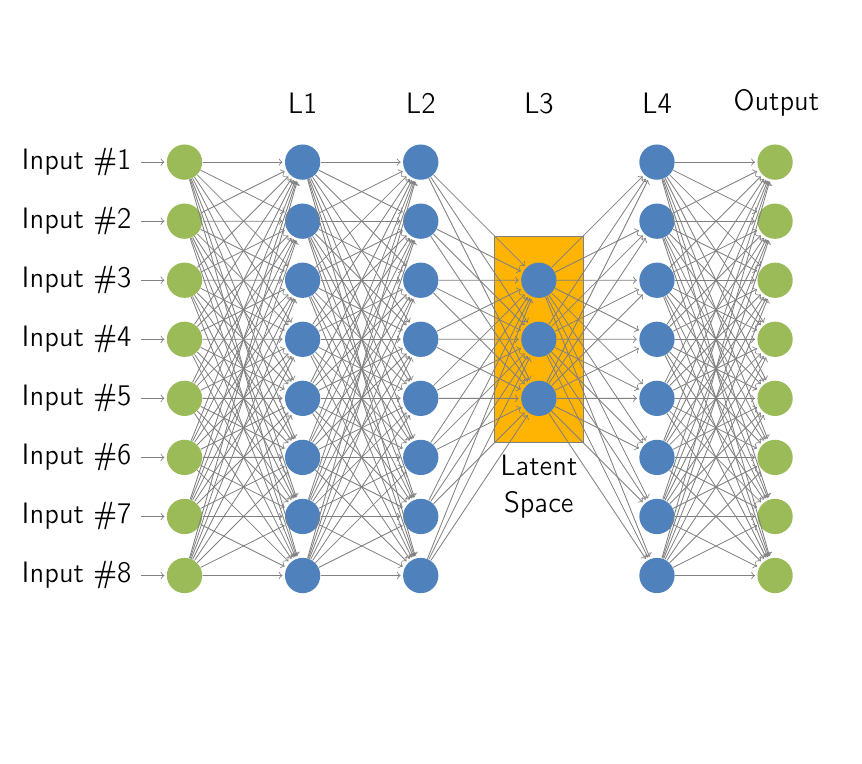}\hspace{0.3cm}
\includegraphics[width=0.47\linewidth, trim={0 0 0 0}, clip=true]{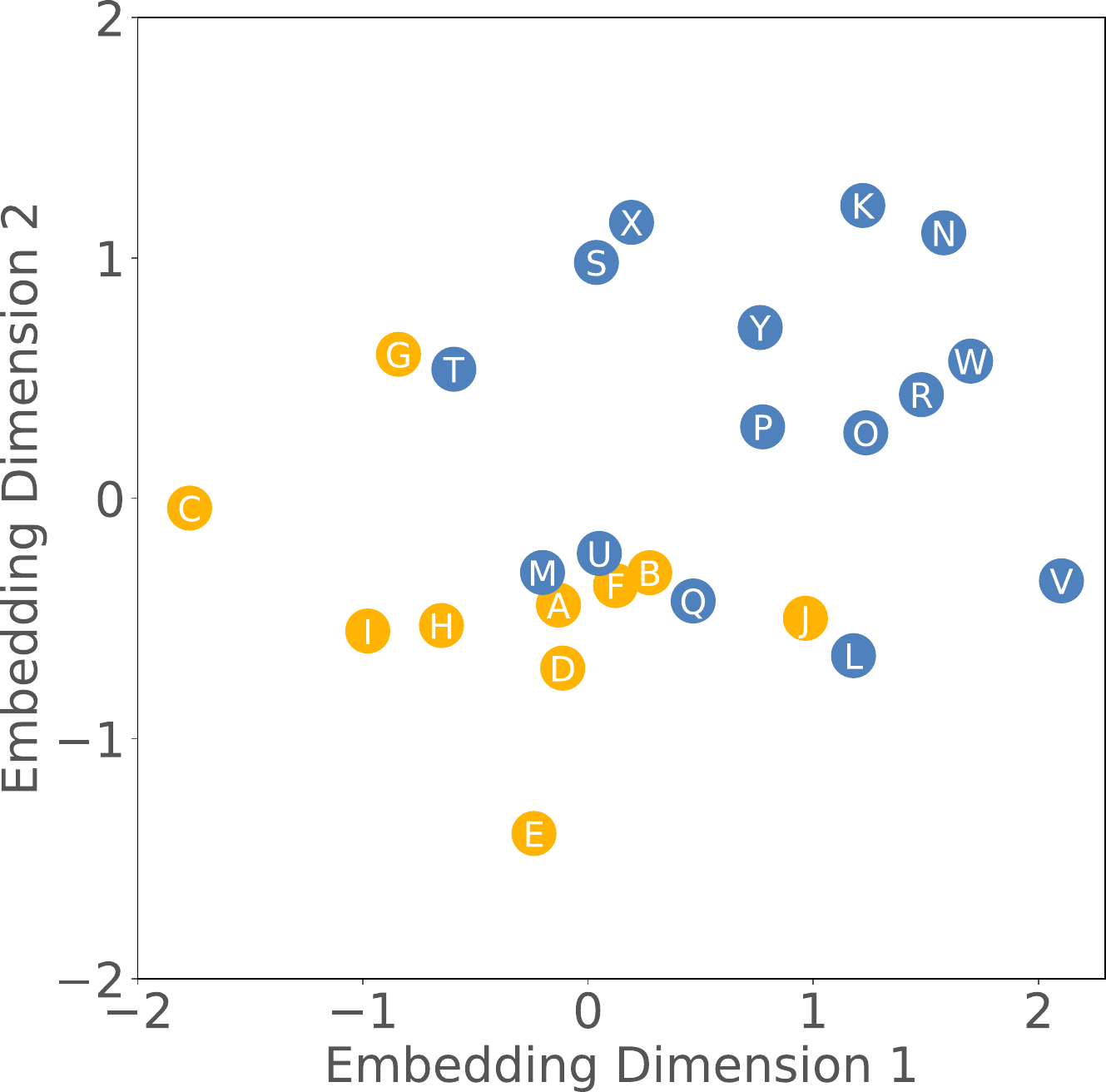}
\caption{A neural network maps input data into an a latent space, and embedding space, respectively. (Left) An autoencoder-like structure in which the network architecture suggests a latent space in Layer 3 (L3), that us commonly referred to as a ``bottleneck''. Autoencoders are often trained by supplying a training target equal to the input data, so that the loss function encourages a latent space to capture features of the inputs. Latent representations can, in fact, be distributed through more than one layer, and can also be referred to as providing an embedding of the data. (Right) An illustration of how a well-trained network might separate data in a 2D latent space, or a 2D subset of latent space for a task of diagnosis. Here, data points used in training, corresponding to individual patients (lettered), are assigned class labels (yellow, or blue, indicating diagnostic outcome). This data distribution is synthetic, according to an assumed bimodal, bivariate Gaussian mixture model, but embeddings of this nature can be learned for patient datasets containing multiple diagnoses \cite{chushig2022learning}.}
\label{fig:EmbSpace}
\end{figure}

For a diagnostic ``task'', we map data from new patients into an embedding space, and use, perhaps, a simple rule to determine whether the patient has the condition based on comparing the location of the new (unlabelled) data to the (presumably gold-standard labelled) embedding that has already been learned. For example, a nearest neighbour decision rule would be one obvious choice.

But, using the nearest neighbour decision rule, a membership inference attack could be performed. Within the example embedding illustrated in Figure~\ref{fig:EmbSpace}, we may visualise the nature of a potential privacy violation (see Fig.~\ref{fig:EmbSpaceWithQuery}), by imagining a database of points from patient measurements or observations with diagnostic ground truth mapped to this embedding space.

The conceptual illustration might seem to suggest an unlikely scenario: we have access to the internal representation of a deep network. However, this is only for the purposes of illustration; it turns out that in having access to the outputs of a network, we can, under some circumstances, figure out whether a patient record, which we have access to as attackers, was used to train the network or not. This is considered an unintended privacy violation.

\subsection{Membership Inference Attacks}
Assume that there exists a machine learning model that has been trained for some task. We consider the case of a specific data record that has plausibly been included in training the model, i.e. the record contains information that can be consumed by the model in some form. The purpose of a membership inference attack is to determine if a specific data record was used in the training data of the target model. There are two broad classes of MIA, white-box attack scenarios in which the attacker has information about the model's structure and parameters, and black-box scenarios, considerably more tricky, in which one has access only to the outputs of the model, and the ability to provide specific inputs for the inference process of a model treated only as a black-box.

The feasibility of black-box attacks was somewhat surprisingly demonstrated by Shokri \textit{et al} \cite{Shokri2017MembershipModels} in an approach involving the training of so-called \emph{shadow models}. The shadow models do not need to be trained on the precise dataset used to train the target model to be attacked, but should have access to data from the same distribution. The attacker need know only the structure of the input data records, and to have access to the prediction vector for, say, a multi-class decision problem. The attack model is simply taught to observe the output of a collection of networks, known as the shadow models, which are trained with specific data items drawn from a similar population and record structure to that of the target model. The attack model -- with access to the included and omitted training data items -- learns to detect the differences in the shadow models' responses, during shadow inference, to data items that might have been included or excluded from the training data for each shadow model.

As a general rule, attack models perform better when there is more information available about the nature of the training of the target model, the distribution of data used to train it, and the architecture of the target network. Several studies have subsequently extended the shadow attack methodology, improved scalability \cite{bertran2024scalable}, or demonstrated application to other types of inference problem such as regression \cite{gupta2021membership}.  

Membership inference attacks might be successful due to the fact that models often overfit to the training data, leading to a response that indicates higher confidence at inference time for input data that have been seen during training. Although early membership inference attacks on deep networks looked explicitly at prediction confidence, 
\citet{chen2013expressive} demonstrated that even with label-only access to the target network's output, information can be gleaned about the structure of the decision boundary around a data item through repeated queries to the target model in which perturbations are introduced, and label shifts observed. This again provides a sufficiently strong queue to an attack model about data item membership during training. For a complete survey of techniques, see~\citet{hu2022membership}.

\subsection{Differential Privacy}
\label{subsec:DPIntro}
Though there are several variants, differential privacy seeks to place an upper bound on the maximum change in distribution of some quantity of interest that is captured by the data under some change to a dataset.

Let $x$ and $x^{\prime}$ be two collections of records (datasets) that differ by a single entry, and where one entry corresponds to the record of one specific individual within the dataset. Then, we consider a mechanism, $\mathcal{M}$ that maps collection $x$ or $x^{\prime}$ to an output, $s$, (which could be a query output) about the collection. The mechanism is said to be $\epsilon$-differentially private if
\begin{equation}
   \log_e \left ( \frac{Pr \left [ \mathcal{M}(x) = s \right ]}{Pr \left [ \mathcal{M}(x^{\prime}) = s\right ]} \right ) \leq \epsilon, \forall s \in S 
   \label{eq21}
\end{equation}
where $S$ represents the space of possible outcomes of the mapping $\mathcal{M}$; in this definition, the probabilities are intended to be over the space of a stochastic definition of $\mathcal{M}$, so that multiple queries on the same datasets return slightly different results. 

\begin{figure}[ht!]
\centering
\includegraphics[width=10cm]{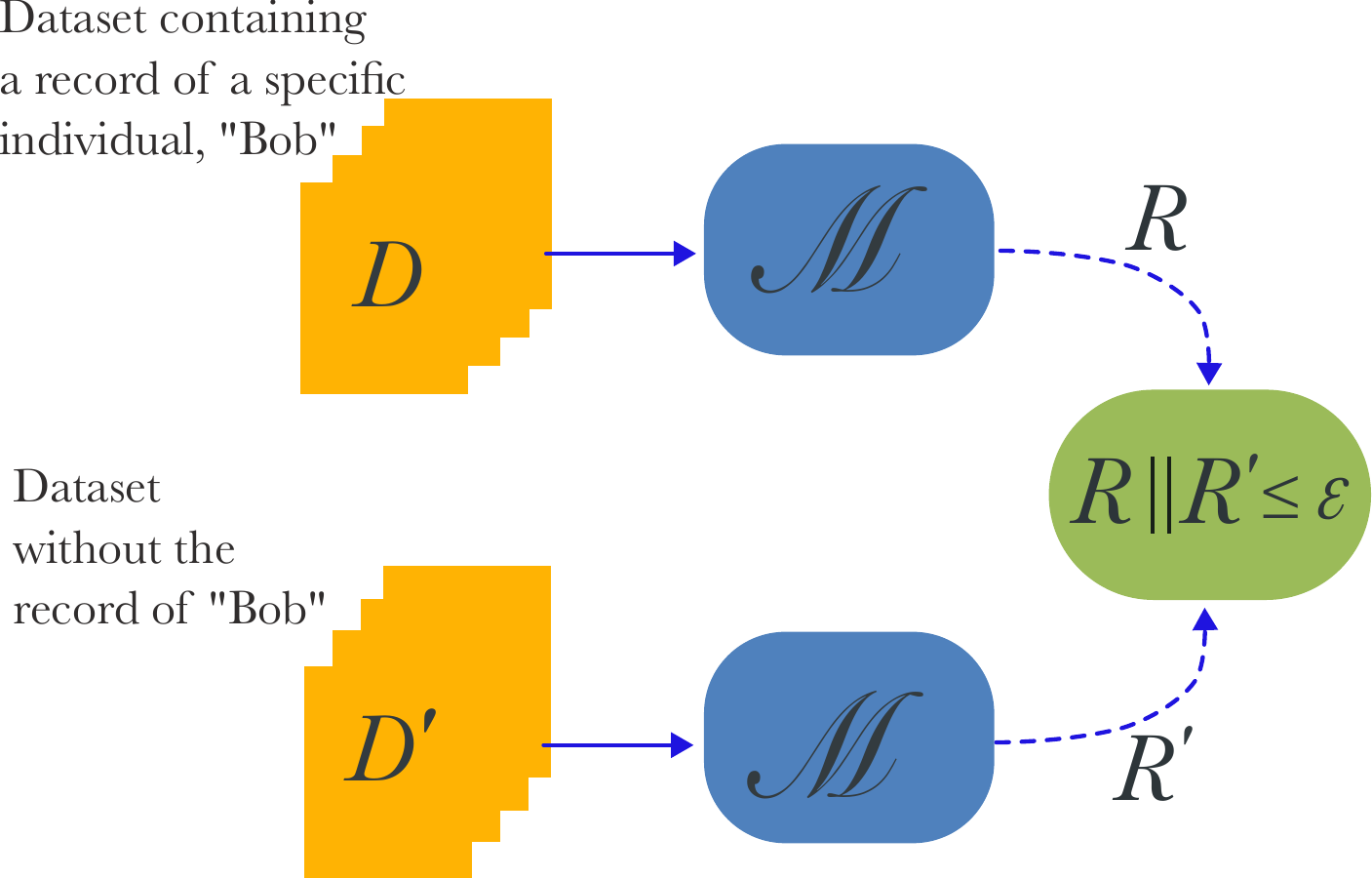}
\caption{A simplified illustration of the principle of differential privacy. Two datasets, one containing $(D)$ and one omitting ($D^{\prime})$ the record of a particular individual (``Bob'') are processed by a mechanism, $\mathcal{M}$ to protect the privacy of the data. Subsequent summarising queries, $S_Q$, generate similar responses on both datasets. This is captured by computing or measuring the ratios of probabilities across the possible responses, $R$, from the queries, and establishing bounds on the log-ratios of probabilities $R||R^{\prime}$ across response space from $D$ and $D^{\prime}$.}
\end{figure}

Some remarks about the definition of Eq~(\ref{eq21}) are in order; first, this defines $\epsilon$ differential privacy. A relaxed definition due to~\citet{dwork2014algorithmic}, is known as \emph{approximate} $(\epsilon, \delta)$-privacy, where the mechanism is allowed to ``fail'' with probability $\delta$ and no guarantees are provided. While this sounds like a grave violation of the principle, by choosing small enough $\delta$ and with theoretical guarantees of how algorithms implementing such approximate differential privacy behave wrt. to $\delta$\footnote{More specifically, ``no privacy guarantees at all'', would be a possible mechanism failure. In practice, relevant implementations fail ``graciously'', in that the probability of leaking increasing private information decays quickly~\cite{Mironov2017RenyiPrivacy} (even sub-Gaussian~\cite{bun2016concentrated})}, $(\epsilon, \delta)$-privacy makes for a more manageable privacy , as it allows for a wider range of ways to add noise.  
Secondly, the nature of the change to the dataset between $x$ and $x^{\prime}$ is quite vague. The change might consist of removal of a record, or modification of the entry associated with one individual. In later discussions, we will focus on the \textit{removal} of a record, as this is tied to a well-established form of attack on privacy, known as a membership inference attack (MIA).

The importance of definition Equation~(\ref{eq21}) is is that it places a bound on the change in distributions of some summarisation of the data given a change to the data itself. Specifically, given a privacy mechanism, and a description of the two datasets $D$ and $D^{\prime}$, which differ in a certain record, we can link the privacy mechanism to an upper bound on the ratio described on the left hand side of Equation~(\ref{eq21}). We will illustrate the nature and origin of this bound by considering how it is computed for a specific example. To do this, we must consider a question or characteristic of the data that would be captured by $\mathcal{M}$, the mechanism that we wish to apply to the two ``versions'' of the dataset, $D$. Ignoring any randomisation (equivalent to $\epsilon=\infty$), common summarising statistics such as mean or counts.

\begin{figure}[ht!]
\centering
\includegraphics[width=0.47\linewidth, trim={0 0 0 0}, clip=true]{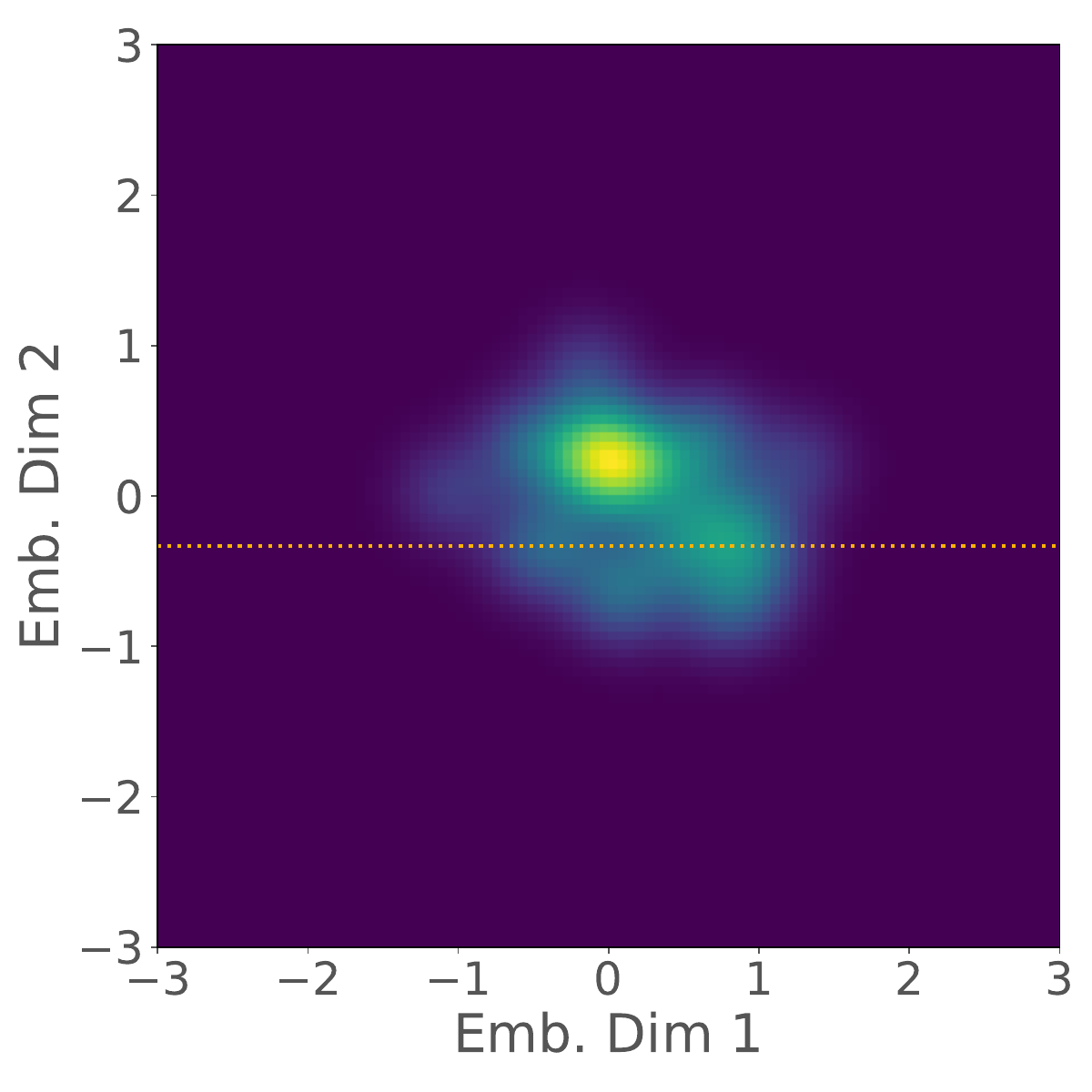}
\hspace{0.3cm}
\includegraphics[width=0.47\linewidth, trim={0 0 0 0}, clip=true]{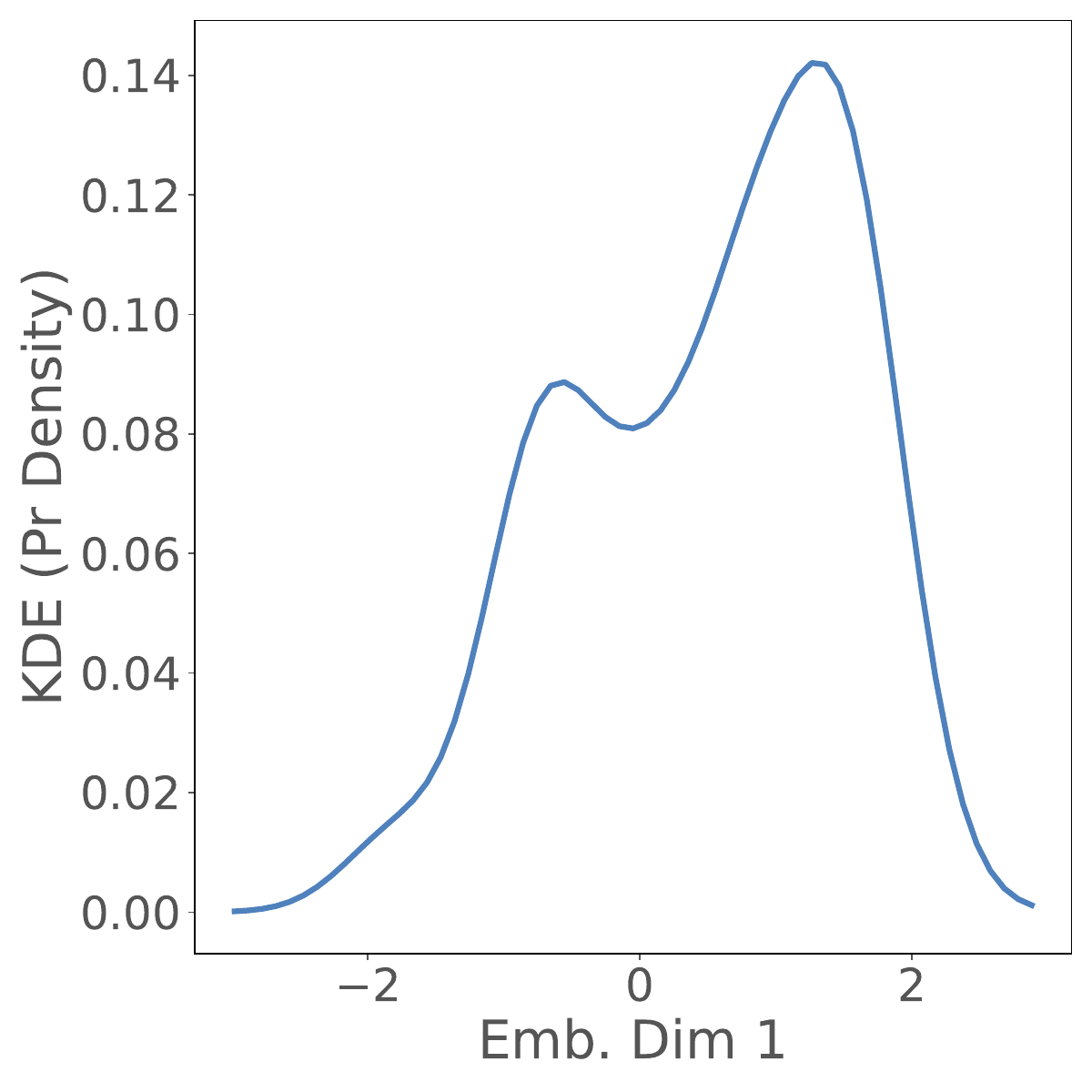}
\caption{(Left) A kernel density estimate of the toy embedding problem from the embedding space example of Figure~\ref{fig:EmbSpace}; line overlay indicates position of profile shown on right. (Right) a profile through the middle of the embedding space; we can deduce the presence of underlying clusters, even though they are not precisely attributable to the underlying classes.}
\label{fig:KDE}
\end{figure}

Given the importance of deep learning, we instead consider an estimator of the total distribution of embedded patient data, a more descriptive summary of the distribution of data points. Consider a kernel density estimate of the distribution of data in embedding space calculated by:
\begin{equation}
\hat{p}_D(\mathbf{x}|h) = \frac{1}{N \cdot h}\sum_{n=1}^N K \left (\frac{\mathbf{x}- \mathbf{x}_n}{h} \right )
\label{eq:KDE}
\end{equation}
A kernel density estimate of the toy embedding problem shown in Figure~\ref{fig:EmbSpace} is shown is presented in Figure~\ref{fig:KDE}.

This definition of differential privacy has some attractive properties. In particular, it admits several relatively simple stochastic algorithms that establish privacy upon a database. Also, ``down-stream'' processing applied to data that has been made $\epsilon$-differentially private is guaranteed to be at least $\epsilon$-differentially private (\emph{post-processing} principle), bounding the likelihood of extracting sensitive information.

\begin{figure}[t]
\centering
\includegraphics[width=0.47\linewidth, trim={0 0 0 0}, clip=true]{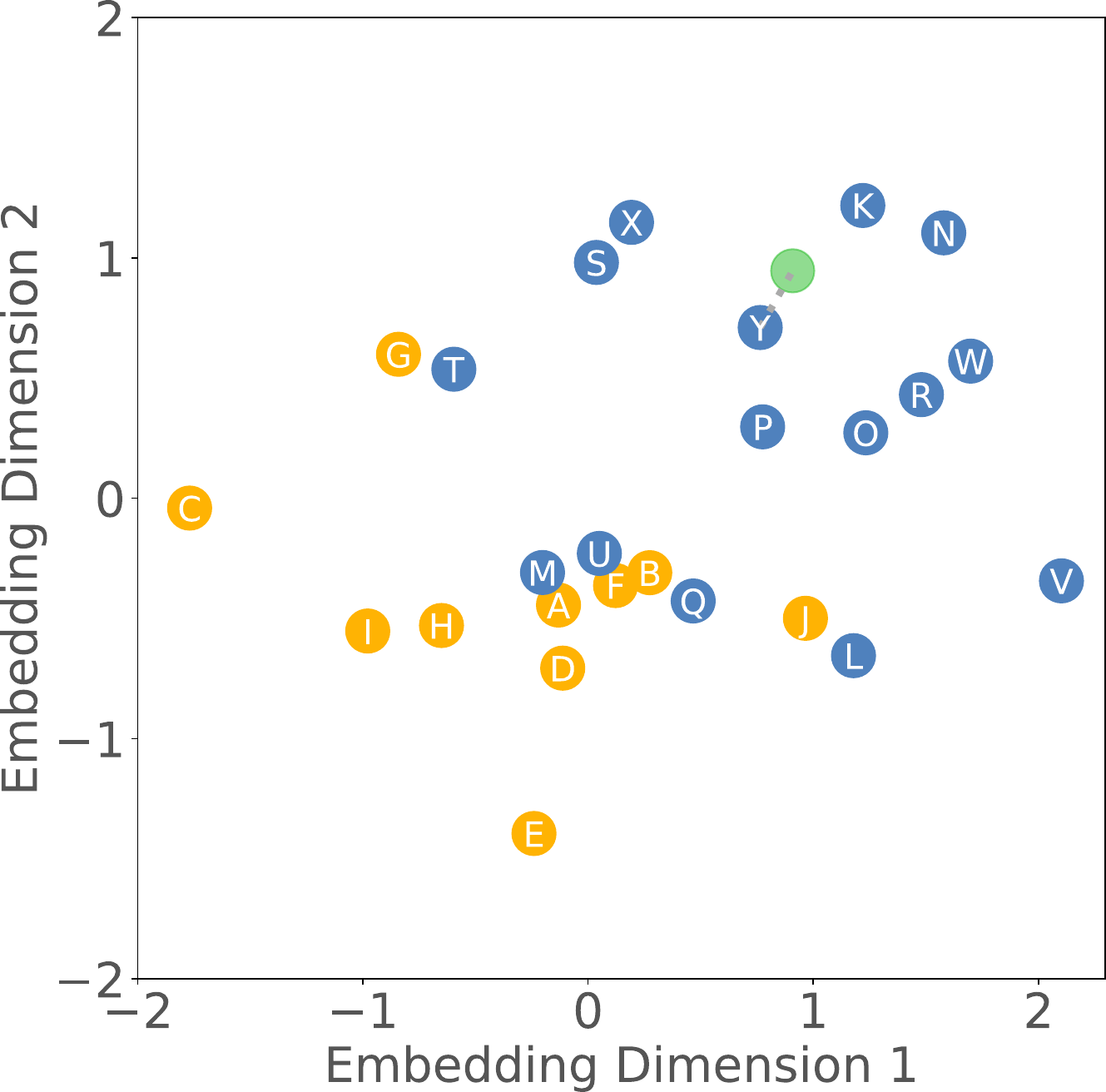}\hspace{0.2cm}
\includegraphics[width=0.47\linewidth, trim={0 0 0 0}, clip=true]{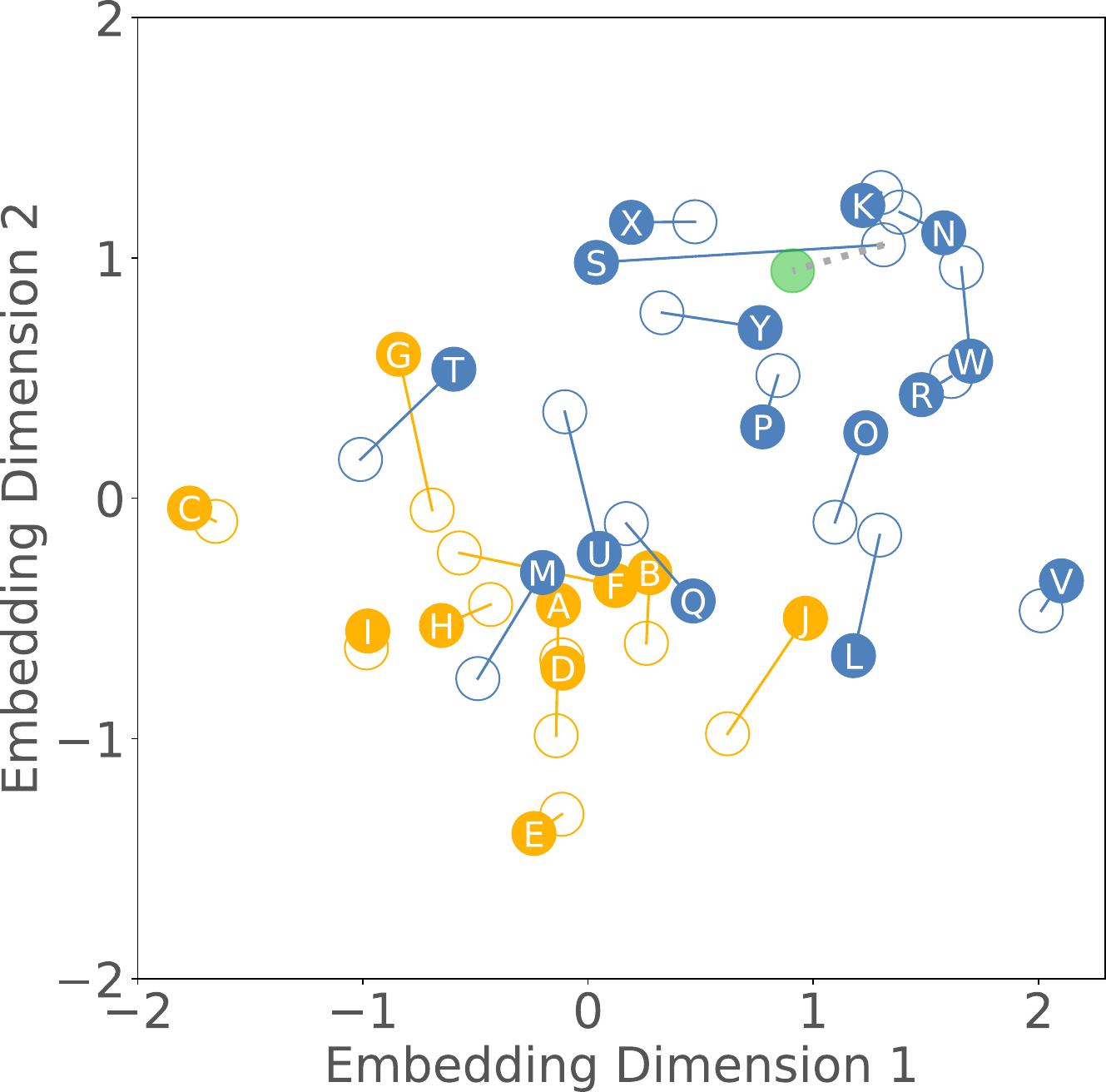}
\caption{(Left) Illustration of associating a query, in which the record of a specific patient is encoded to a representation space, and maps to a location, shown in green. The proximity of this query to the embeddings of (lettered) individuals can be inferred by membership inference attacks through a number of approaches. For example, assume that diagnostic classification is based on applying the nearest (Euclidean sense) neighbour rule: in this example, the diagnosis would be quite certain, but so is the identity, by the same rule! (Right) Addition of noise to the embedding reduces the chance of re-identification, and (ideally) leaves diagnostic classification largely intact: the ideal outcome for a privacy mechanism. However, this example represents a particular realisation of noise that is illustrative of the principle of privacy preservation. Multiple queries can still reveal likely membership, and useful class distinction is seldom this well preserved.}
\label{fig:EmbSpaceWithQuery}
\vspace{-1em}
\end{figure}
\begin{wrapfigure}{R}{0.50\linewidth}
\vspace{-3.1em}
\centering
\includegraphics[width=0.95\linewidth, trim={0 0 0 0}, clip=true]{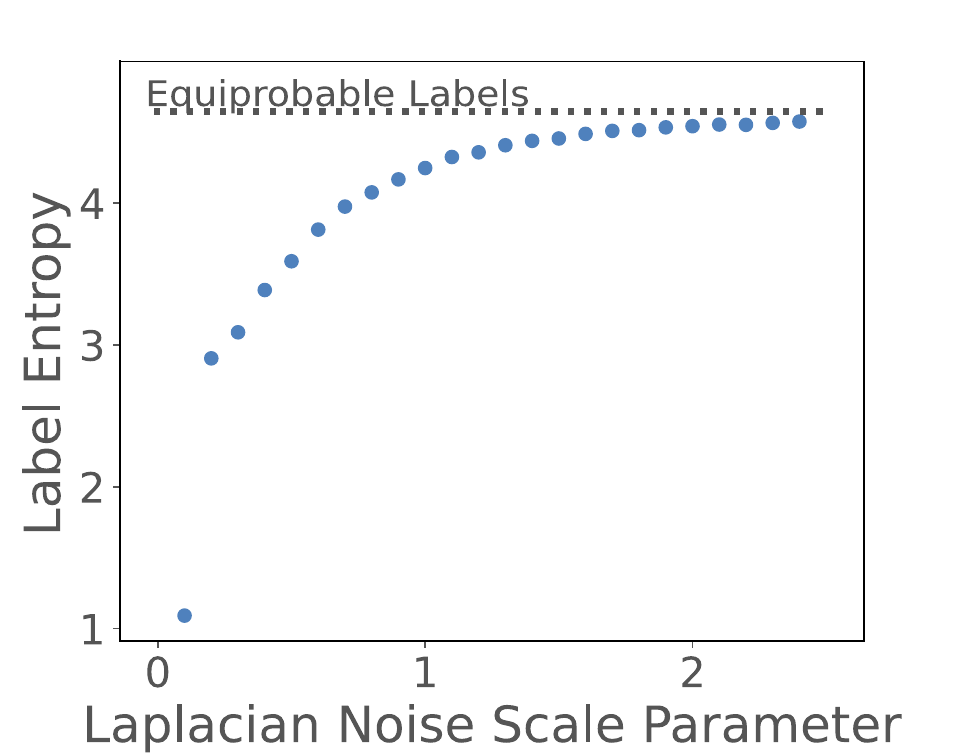}
\caption{A toy membership inference attack based on repeated nearest neighbour querying given the embedding of a target becomes more difficult with increasing quantities of noise. This is reflected in the label entropy in response to repeated NN queries. As noise is increased, all database items are equally likely to be returned in response to the query target, protecting against membership inference.}
\label{fig:NoiseProtection}
\vspace{-2.1em}
\end{wrapfigure}
\sloppy
\subsection{Example: Privacy and Embedding Spaces}
Differential privacy can be applied to many different forms of data and at several stages of its use, including during the learning process. 

In Fig.~\ref{fig:EmbSpaceWithQuery}~(Left), the space represents a \textit{pair} of real-valued functions. We can look at the simple membership inference problem described by a nearest neighbour approach and consider the effect of adding a stochastic perturbation to the embedding. Adding Laplacian noise was suggested as one mechanism to achieve differential privacy~\cite{dwork2014algorithmic}. The effect is trivially shown in Fig.~\ref{fig:EmbSpaceWithQuery}~(Right), where the nearest neighbour is now no longer a (very similar) point from the ground-truth data, but instead that of another individual. In this particular noise realisation, an attack picks the wrong individual. However, one really needs to assess the probability of returning the identity of a target individual over multiple realisations of the (stochastic) privacy mechanism $M$, and to assess to what extent a simple histogram of results from queries might reveal a significant preference for a certain individual. This is somewhat captured in measuring the entropy of label distribution in response to multiple queries (see Figure~\ref{fig:NoiseProtection}).

\vspace{2em}
\subsection{Visualising Privacy Bounds}
\begin{figure}[ht!]
\centering
\includegraphics[width=0.48\linewidth, height=0.48\linewidth, trim={0 0cm 0 0cm}, clip=true]{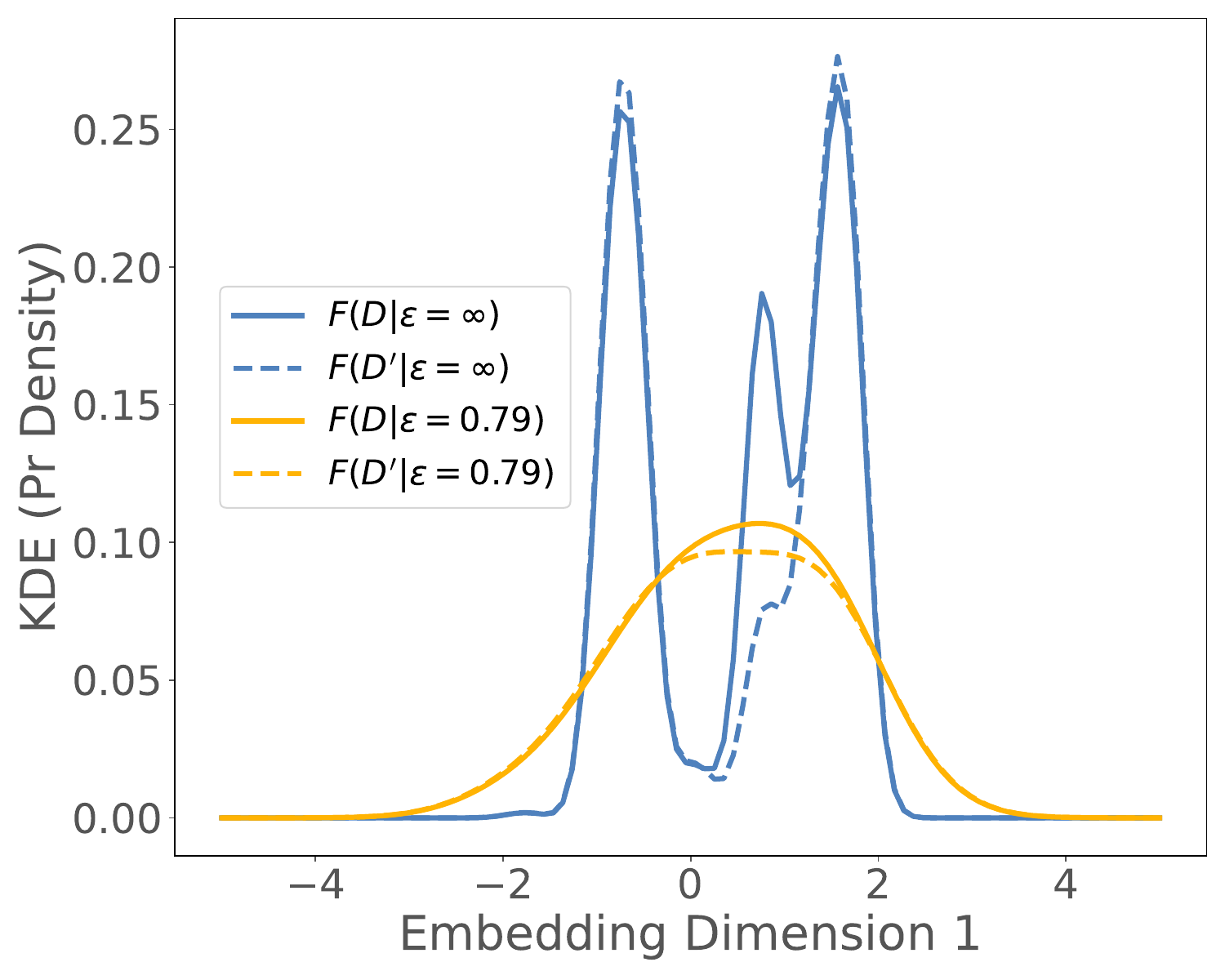}
\includegraphics[width=0.48\linewidth, trim={0 0 0 0}, clip=true]{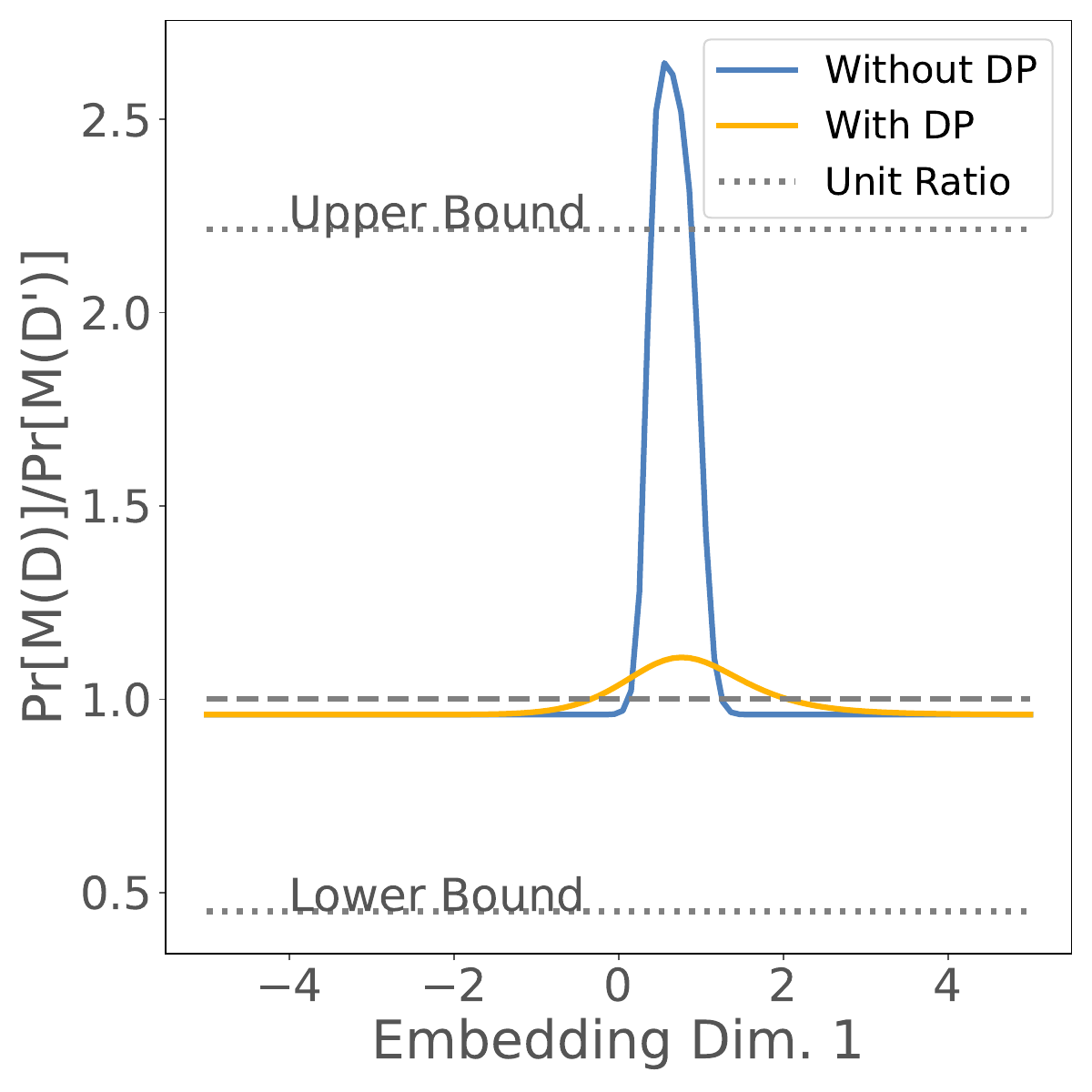}
\caption{(Left) A comparison of profiles through KDE estimates of two versions of a dataset, one with 26 embedded patient records, the other with 25. The profile location is the same as that illustrated in Figure~\ref{eq:KDE}; (Right) Ratio of LHS of Equation~(\ref{eq21}), and the privacy bounds. See text and Appendix for details.}
\label{fig:DPProf}
\end{figure}

The bound, $\epsilon$, provided by differential privacy is partly dependent on the maximum change that could be observed on removal of a patient's embedding -- the (maximum possible) impact of any input data point on the observed result is called the \emph{sensitivity} of the mechanism. This has to be considered with care: the application of KDE as indicated by Equation~(\ref{eq:KDE}) will itself partially mask the change introduced by removal of a single patient entry, with the masking effect dependent on the kernel width. We can observe the effect of a Laplacian mechanism~\cite{dwork2014algorithmic} on the profile of KDE estimate with respect to the bounds predicted by differential privacy in Figure~\ref{fig:DPProf}. In this case, $\mathcal{M}$ is in fact a function of location in embedding space, and the ratio of probabilities is estimated by Monte Carlo simulation over 100,000 realisations of the privacy mechanism.

\subsection{Generative models}
Generative models may be succinctly characterised as any algorithm that:
\begin{quotation}
\noindent ...given some samples of data that come from an underlying distribution, $p_W(x)$, creates more samples, $y$, from the same distribution.
\end{quotation}

For a trivial anaology, imagine observing the heights of people in a crowd, estimating the mean and variance of heights from that sample, and (computationally) drawing more samples of the heights of hypothetical individuals from the same population under a Gaussian approximation of the height distribution. In modern generative modelling, the underlying data distribution is typically multivariate, multimodal (in the statistical sense), and implicitly captured by the parameters $\mathbf{\theta}$ of a deep neural network. In practice, the generated data is actually drawn from a distribution that describes the parameterised data-generating model, $p_G(y;\theta)$, and one hopes that $p_G \approx p_W$.

The earliest examples of generative models based on deep learning, and trained by backpropagation, include variational autoencoders \cite{kingma2013auto}, which are in some ways closest in spirit to the aforementioned height-of-people example, in that internally, the distribution parameters of a driving noise source for the decoder of the autoencoder are estimated by the encoding process; to draw new samples, we computationally sample from a multivariate Gaussian (easy) and feed those samples into the decoder. However, there have long been schools of thought in both cognitive psychology and Markov-Chain Monte-Carlo techniques that might be described as \textit{analysis by synthesis}, whereby top-down and bottom (data) up computations, \cite{dayan1995helmholtz,hinton1995wake,tu2005image} are alternated to learn the parameters of statistical models from which new samples can be drawn. Even earlier, generative models were used to capture the implicit states of internal networks of simple binary neuronal units known as Boltzmann machines \cite{ackley1985learning}, influenced by observations of external states.

Generative Adversarial Networks (GANs), where two networks are trained in competition, are another example of generative models. The principle behind the generative models of GANs is that, in training the generative model, two networks compete: one deep network, with no access to samples of data, is trained to map samples from an ``easy'' noise distribution into the space of the data observations to be modelled. A second network, the discriminator, is trained to distinguish between samples drawn from the generator network and real observations. The signal used for learning is based on the output of the discriminator and the ``truth'' of whether the sample shown to it has come from the generating network, or are samples of the real data (e.g. real-world observations). Once well-trained, the generator can be used to synthesize new samples.

Diffusion models are one of the most recent and widely deployed of deep generative models, particularly encountered in image synthesis. They arguably arise from the principle of score-matching \cite{hyvarinen2005estimation}. The idea is to reverse the process of adding noise to observations of data. Adding white noise to samples from a multivariate distribution has an equivalence to a diffusion process in the joint probability space of the high-dimensional data. If the diffusion process is time-dependent (as a thermodynamic process might be), then a network can be taught to approximately reverse the process as a function of the time variable; time is only treated as a parameter of the mapping performed by the network, and the reversal of the process of adding noise to a single image is only approximate, but that is in the very nature of generative models.

Numerous other algorithm classes exist, including normalizing flows and generative models based on Gaussian Processes (GPs). Normalizing flows learn a map between one distribution that is easy to sample from, and one that is more difficult. Whilst this typifies nearly all approaches to computational sampling -- including algorithms for Gaussian distributed pseudo-random number generation, such as the Box-M{\"u}ller algorithm \cite{golder1976box}  -- normalizing flows attempt to learn a series of differentiable {\em and} invertible mappings (the flow), all operating on ``proper'', in the sense of being normalised, probability density functions \cite{papamakarios2021normalizing}. The flow maps a base distribution that is easy to sample from to one that is significantly more complex. Normalizing flows have been successfully applied to create image generators.

In contrast, Gaussian Processes (GPs) were initially proposed in the context of time-series modelling; they can be conditioned on functional families with respect to variables such as time or space. Priors may be applied over the possible forms of functional families, imposing strong temporal or spatial structure. Although they are not often thought of in the context of generative models, the fact that their underlying construction involves estimating the parameters of Gaussian distributions over paramaterised functions of (say) time means that they can be used to generate new samples of data. In addition, although generally reported as being computationally expensive, scalability to large datasets has been demonstrated with the right choice of computational architecture \cite{deisenroth2015distributed}.

Finally, borrowed from a computational linguists' toolkit, language models estimate the probability of observing a sequence of tokens  (i.e., words in a sentence), which can be expressed as a Markov chain~\cite{markov1913essai}---the conditional probability of observing a token having observed its predecessors. Sampling from these probability distributions allows to generate new token sequences.
These probabilities were estimated explicitly at first~\cite{shannon1948mathematical}, for a fixed number of tokens~$n$ ($n$-gram language models), but more recently they are estimated with neural networks~\cite{bengio2003neural}. Such neural language models have been shown to yield expressive representations of language~\cite{Peters2018}, and various improvements to their underlying neural network design~\cite{Vaswani2017} that allow to scale both the network size and the amount of training data led to the impressive developments we are witnessing today~\cite{OpenAI2023GPT-4Reportb-abbr}.

\section{Methods for generating synthetic data}
\label{sec:methods}
Overall, methods for generating synthetic datasets with privacy guarantees can be divided into three main approaches. 
Firstly, private datasets can be directly perturbed, e.g., by projecting them into a low-dimensional vector space, adding DP-satisfying noise to the projections and then projecting back to the original input space~\cite{Gondara2020DifferentiallyProjections}. 
While this provides a quick solution for small datasets, this method does not scale well to larger datasets due to increasing sensitivity of the method and thus the arising requirement to add more noise. 
It is also not clear how these methods would apply to datasets that require discretised inputs (i.e., as is the case with textual tokens). 
Secondly, private datasets can be used to train generative models, such as Generative Adversarial Networks ~\cite{Goodfellow2014GenerativeNets}, Variational AutoEncoders~\cite{kingma2019introduction}, Normalizing Flow~\cite{papamakarios2021normalizing},  Diffusion Models~\cite{Rombach2022High-resolutionModels} or language models~\cite{Schick2021GeneratingModels}, while satisfying DP guarantees during model training, e.g. by clipping gradients to bound the sensitivity of training and adding noise to them, thereby obfuscating the effect of each training sample, as proposed by~\citeauthor{Abadi2016DeepPrivacy}~\cite{Abadi2016DeepPrivacy}. 
Via the post-processing property, the data generated by models trained in this way is subject to the same privacy guarantees as the model. 

Finally, another group of approaches seeks to explicitly align the distributions of the generated data with that of the private datasets~\cite{Zhao2024GeneratePrivacy}.  DP is achieved by modifying the comparison mechanism rather than the training procedure itself which by design of the algorithm can reduce the required amount of noise~\cite{Harder2021DP-MERF:Generation,Lin2024DifferentiallyImages}.


\subsection{Tabular/Structured data}

The literature on differentially private synthetic data generation has traditionally focused heavily on structured (tabular) data formats. While a comprehensive review of methods is beyond our scope (we refer readers to recent surveys such as the one from \citet{Hu2024SoK:Synthesis}) 
we highlight some of the major developments. 

Statistical methods represent one key direction, including marginal-based methods that model correlations between attributes, for example using Markov Random Fields (MRF), where nodes represent attributes and edges capture pairwise dependencies. 
The MRF-based methods typically construct a graph exploring pairwise dependencies between attributes, applying junction tree algorithms to obtain the MRF, from which noisy marginals are generated for synthetic data sampling in order to satisfy DP guarantees. 
A notable example includes PrivBayes~\cite{Zhang2017PrivNetworks} which uses Bayesian networks to model attribute relationships. 
Specifically, PrivBayes constructs a differentially private Bayesian network by first privately learning the network structure through selecting parent-child relationships between attributes using 
mutual information as the score function. 
It then injects Laplace noise into the conditional distributions associated with each edge in the network before using these noisy distributions to generate synthetic data. 

Copula-based approaches provide another powerful framework by separating the modelling of univariate marginal distributions from their dependency structure using copula functions~\cite{li2014differentially}. 
These methods first estimate univariate margins and the correlation matrix privately, then derive a Gaussian or vine copula to capture complex dependencies between attributes. 
Differential privacy is typically achieved by adding calibrated Gaussian or Laplace noise to the estimated correlation matrix (in the case of Gaussian copulas)~\cite{Asghar2019DifferentiallyCopula} or to the parameters of the selected bivariate copulas (for vine copulas)~\cite{gambs2021growing}, with the noise scale determined by the sensitivity of these parameters and the desired privacy budget. 
An alternative direction that avoids explicitly learning probability distributions is provided by projection-based approaches---these methods work by projecting the original dataset into a lower-dimensional space using techniques like randomised projections~\cite{Dwork2014AnalyzeAnalysis}, adding noise in this compressed representation, and then reconstructing synthetic data~\cite{Gondara2020DifferentiallyProjections}. 
This can be an effective strategy for smaller-scale datasets where direct perturbation is feasible.

Deep generative models form another important category, particularly various privacy-preserving GAN enhancements---indeed, the proliferation of DP-GAN variants has spawned dedicated surveys on this subtopic alone~\cite{Fan2020ANetworks,10.1145/3459992}.
A seminal approach is DP-GAN~\cite{Xie2018DifferentiallyNetwork}, which achieves differential privacy by injecting noise to the discriminator's gradient during training. 
While the discriminator has access to both real private data and generated samples to perform its classification task, the estimation of the generators' parameters only sees the noisy gradients from the discriminator rather than the raw private data. Therefore, by the post-processing property of differential privacy, the generator's outputs also satisfy differential privacy, and, subsequently, the data generated by the generator. CTAB-GAN+~\cite{zhao2024ctab} adopts the DP training strategy from DP-GAN and applies it specifically to tabular data synthesis. 
PATE-GAN~\cite{Jordon2018PATE-GAN:Guarantees} takes a different approach by building on the Private Aggregation of Teacher Ensembles (PATE) framework~\cite{Papernot2017Semi-supervisedData,Papernot2018ScalablePATE}, where an ensemble of teacher models is trained on disjoint subsets of private data, and their aggregate predictions (made differentially private through a noisy voting mechanism) are used to label public data for training a student model.
In PATE-GAN, multiple teacher discriminators are trained on disjoint subsets of the private data. Rather than requiring public data for training the student discriminator, PATE-GAN uses samples from its own generator as training data. These generated samples are labelled as ``real'' or ``fake'' through a noisy voting mechanism that aggregates the teachers' predictions, providing differentially private supervision to the student discriminator. Since the student discriminator is trained only on generated samples and noisy aggregated labels, and the generator in turn only interacts with this student discriminator, differential privacy is preserved through the post-processing property.

An emerging direction in tabular generation leverages the robust pre-training of large language models to enhance the privacy‑utility trade‑off. Pre-trained models inherently capture rich, general knowledge about language and data structures (expressed in language), such as table schemas and inter‑attribute relationships, without incurring any privacy cost. By partitioning the learning process, these methods enable the model to first acquire non‑sensitive structural information from public or less sensitive data and then concentrate privacy‑preserving measures exclusively on sensitive data during fine-tuning. For example, the DP‑2Stage framework \cite{dp2stage} initially fine-tunes a GPT‑type LLM on a synthetic template table derived from public statistics so that the model learns the overall table structure without noise. It subsequently applies DP‑SGD solely to learn the sensitive cell values in the private dataset, which leads to substantial down-stream utility improvements 
In a complementary approach, the HARMONIC architecture \cite{harmonic} uses instruction fine-tuning combined with a $k$‑nearest neighbour based retrieval mechanism to capture common data formats. This architecture computes the training loss solely on the synthetic output, which minimises the risk of memorising exact sensitive values while still benefiting from structural learning. 
Moreover, parameter‑efficient fine-tuning techniques such as LoRA‑based tuning \cite{dplora} build on the pre-trained model by freezing nearly all its weights and updating only a small set of additional parameters in a low‑rank subspace. Although the empirical evidence for an improved privacy‑utility trade‑off using LoRA‑based tuning remains preliminary, the intention is to capitalise on the robust pre-trained representations while restricting the introduction of differential privacy noise to a compact subset of parameters. 


\subsection{Images}

Building on the success of GANs for tabular data, several works have explored differentially private image generation. 
However, achieving high-quality image synthesis while maintaining meaningful privacy guarantees has proven challenging, especially for complex natural images.
Early approaches like DP-GAN~\cite{Xie2017} and DP-CGAN~\cite{Torkzadehmahani2019DP-CGAN:Generation}, an extension of the GAN architecture to allow generation conditioned on additional information (typically class labels), primarily focussed on simpler datasets like MNIST and FashionMNIST~\cite{lecun1998mnist,Xiao2017Fashion-MNIST:Algorithms}. 
Even with relatively weak privacy guarantees ($\epsilon \approx 10$), these methods struggled to capture the full complexity of the data distributions, often resulting in blurry or mode-collapsed samples.
The fundamental challenge stems from the noise required for DP-SGD training of the discriminator, which increases with the model's dimensionality and makes training larger architectures needed for complex images less practical~\cite{Harder2023Pre-trainedGeneration}. Another possibility is that the amount of noise added during SGD optimisation is overestimated by the privacy accountants (the method to estimate the amount of noise to add to each mini-batch gradient, to maintain the guaranteed ($\epsilon$, $\delta$)-DP inequality), as better compositionality theorems have been proposed specifically for the Gaussian noise mechanism~\cite{Dong2022GaussianPrivacy}.

More recent work has made progress by leveraging auxiliary public data.
DP-MEPF~\cite{Harder2023Pre-trainedGeneration} demonstrates that pre-trained perceptual features from public datasets can enable high-quality private generation of more complex datasets like CIFAR-10 and CelebA, even under stronger privacy constraints ($\epsilon \leq 2$). 
Rather than training a GAN adversarially, DP-MEPF minimizes the difference between the distribution of private data and and the output distribution of a GAN as measured by Maximum Mean Discrepancy (MMD). 
This allows privatizing the data-dependent term just once rather than adding noise at each training step as in DP-SGD. 
The use of pre-trained features is an improvement over the initial DP-MERF framework~\cite{Harder2021DP-MERF:Generation}, which relies on random perceptual features instead of pre-trained ones.
GANobfuscator~\cite{Xu2019GANobfuscator:Privacy} and other recent approaches~\cite{Fan2020ANetworks} have also explored techniques like adaptive gradient clipping and warm-starting with public data to further improve training stability and sample quality.
However, generating high-resolution natural images with meaningful privacy guarantees remains an open challenge. As noted by~\citeauthor{Harder2023Pre-trainedGeneration}~\cite{Harder2023Pre-trainedGeneration}, there is still a significant gap between private and non-private deep generative models, particularly for complex high-dimensional data distributions, such as images.

Beyond GANs, researchers have explored other generative architectures for differentially private image synthesis. Recent work has shown promising results using diffusion models~\cite{Ghalebikesabi2023DifferentiallyImages}, discussed in Section~\ref{sec:background}. 
By leveraging ImageNet pre-training and using DP-SGD during fine-tuning, DP-Diffusion achieves state-of-the-art results on complex datasets like CIFAR-10, producing realistic images under relatively weak guarantees. They demonstrate that diffusion models can effectively handle the added noise from differential privacy. 
Proposed more recently, the Private Evolution algorithm~\cite{Lin2024DifferentiallyImages} takes a different approach by treating generative foundation models as black boxes only accessible via APIs and aligning the distribution of an initially random set of generated images to that of the private data. The alignment is done iteratively, by first computing a differentially private histogram of which generated images are most similar to each private image, then generating variations of the most similar images while discarding the least similar ones. This allows leveraging powerful pre-trained models without requiring access to their weights or architecture details. Despite this more restrictive way of accessing generative models, they achieve state-of-the-art results on multiple benchmarks, in part due to the power of pretrained image generators but also due to the fact that noise is being added to the histogram rather than to parameter updates. Because histograms are built by counting private image votes, the sensitivity (i.e. the contribution of each private data point to the overall result) of the algorithm is low, therefore less noise is required to obfuscate each private sample's contribution. 

\subsection{Text}

Traditionally, deep generative approaches for text have lagged behind their continuous counterparts\footnote{e.g., GANs~\cite{Goodfellow2014GenerativeNets} were proposed five years before transformer-based causal language models such as GPT-2~\cite{radford2019language}}. GANs~\cite{Goodfellow2014GenerativeNets}, popular for image synthesis, have not found wide-spread use in text generation, due to the non-differentiality of the argmax operation required to obtain discrete text tokens from the output distribution. Nonetheless, with the proliferation of transformer-based generative language models~\cite{Peters2018,radford2019language}, and their continued success due to the parallelisability and capability increases with parameter count (a form of scalability), there is growth in the number of methods for generating synthetic textual datasets.

A sizable proportion of techniques use a combination of differentially private fine-tuning of (Large) Language Models on private data~\cite{Yu2021DifferentiallyModelsb} with ``prompt engineering''~\cite{Ouyang2022TrainingFeedback}---the generation of text conditioned on specific textual input ``prompts''. Here, models are fine-tuned to re-generate the original text instances of a dataset, given ``control codes'', such as text labels or dataset meta-information, as input. Through the post-processing property, any outputs generated by the fine-tuned model inherit the DP guarantees of the fine-tuning process. Synthetic data can then be generated by prompting the fine-tuned model with the corresponding control codes~\cite{Yue2023SyntheticRecipe,Mattern2022DifferentiallySharing}. It is worth noting that the control codes (and their distribution) are assumed to be public information. This formulation is naturally suitable for single-class text classification datasets, using the class labels as control codes. Indeed, most efforts toward generating synthetic textual datasets are concerned with text classification (e.g., sentiment analysis on reviews) as the down-stream task of the synthetic data.

More recently, methods that aim to align the distributions of generated and private data have also been applied to textual data \cite{Zhao2024GeneratePrivacy}. For example, the Private Evolution algorithm~\cite{Xie2024DifferentiallyText} has been shown to generate high-quality data, by prompting LLM APIs in a black-box manner. However, because unlike image generation APIs, LLM APIs usually do not provide an intuitive function to vary inputs, the variation step is performed by prompting the language model to paraphrase the input. To introduce diversity while maintaining domain coherence (e.g., a tweet might be paraphrased in a different way from a scientific paper abstract), the method relies on manual prompt engineering~\cite{Xie2024DifferentiallyText}.   





\section{Evaluation of Synthetic Data}
\label{sec:evaluation}
The evaluation of privacy-preserving synthetic data typically considers two key aspects: utility, and privacy. Utility can be further broken down into how \emph{realistic} the generated data looks like, compared to the real data, and how \emph{useful} the data can be for down-stream tasks. These aspects are analysed as trade-offs, where utility metrics are evaluated at specific privacy levels determined by the $\epsilon$ parameter of the formal privacy guarantees \cite{Zhang2017PrivNetworks,Jordon2018PATE-GAN:Guarantees,Mattern2022DifferentiallySharing}. 
As privacy bounds tighten (lower $\epsilon$), more noise needs to be added to the generation mechanism, thus degrading the quality of the resulting data.
\subsection{Fidelity}
What we refer to as ``fidelity'' measures how well synthetic data captures the statistical properties and inherent patterns of the original data. This aspect is important where the down-stream applications of the synthetic data might be undetermined and the data is generated to enable exploration and data analysis. 
Intuitively, high-fidelity synthetic data should maintain similar distributions, correlations, and relationships between variables as found in the original dataset. 


For tabular data, fidelity is evaluated in four dimensions---we cover the first three and conclude the section with the fourth as it applies to other modalities as well. First, structural similarity \cite{pdpc2023} ensures that the synthetic dataset preserves the original schema. This involves verifying that every required column appears in the correct order, that data types remain consistent, and that missing value patterns are maintained. For categorical columns, it is crucial to ensure that all expected categories are present because omissions or misrepresentations can introduce bias. For continuous columns, preserving valid value ranges and adhering to formatting conventions such as decimal precision and standardised date formats is vital for maintaining data integrity. Even minor deviations in these aspects can disrupt data processing and lead to analysis errors. Second, univariate distribution similarity assesses whether each feature follows the same distribution in both datasets. For categorical variables, metrics like Jensen–Shannon Divergence \cite{jsd} quantify the overlap between probability distributions, and the category recovery ratio ensures that every category from the original dataset is faithfully reproduced. For continuous variables, KL divergence~\cite{Kullback1951OnSufficiency} or Wasserstein distance \cite{Kantorovich1960MathematicalProduction} measure the optimal transport cost between the distributions, ensuring that the fundamental statistical properties of each variable are maintained. Third, multivariate dependency similarity examines the relationships among features by comparing the correlation matrices \cite{pdpc2023} of the real and synthetic data. This step is critical for applications that rely on complex interactions because it confirms that the joint behaviour and interdependencies among variables are preserved. %

Images and text, being unstructured, elude similar straightforward descriptions. Instead, observed distributions of synthetic and private data can be measured in the embedding space, again, e.g., by means of KL-divergence~\cite{Kullback1951OnSufficiency} or Wasserstein~\cite{Kantorovich1960MathematicalProduction} distance, albeit at the expense of interpretability---a single measure of the distance between two distributions, albeit succinct, arguably carries less explanatory power than the more detailed measures introduced above. As an example, for images, the Fr\'echet Inception Distance (FID) metric \cite{salimans2016improved} is a popular way of approximating the realism and diversity of a synthetic dataset with respect to a real dataset, by comparing the Wasserstein distance of the means and covariances of the embeddings, obtained by the last layer of an Inception network. Similarly, the MAUVE metric~\cite{Pillutla2021MAUVE:Frontiers} measures the distribution divergences between synthetic and real texts by jointly quantising them via $k$-means clustering and measuring the KL divergence of histograms of cluster assignments thus achieved.  Furthermore, for text, focus has also been laid on easy-to-measure characteristics, such as the distribution of text lengths~\cite{Xie2024DifferentiallyText,Yue2023SyntheticRecipe}.

Finally, global structure and diversity of (both structured and unstructured) data can be evaluated evaluated using a combination of propensity score metrics~\cite{tabsyndex} and manifold coverage~ \cite{bellinger2016beyond}. The propensity score method trains a classifier to distinguish real from synthetic data, and an AUC of approximately 0.50 indicates that the classifier is essentially guessing at random, suggesting that the overall high-dimensional distribution is indistinguishable between the two datasets. In parallel, manifold coverage uses dimensionality reduction techniques such as UMAP or t-SNE to compare clustering patterns, confirming that the synthetic data spans the same feature space and captures similar global structure and diversity as the real data. Similarly, for text, topic modelling techniques have been employed to a achieve a similar comparison~\cite{Yue2023SyntheticRecipe}.
 
\subsection{Utility}
Utility evaluation focuses on how well synthetic data performs in specific down-stream tasks. 
This is particularly relevant when the synthetic data is generated for a known application, such as deciding whether a certain document (e.g., patient record) falls in a specific category (e.g., smoker)~\cite{Uzuner2008IdentifyingRecords}.
Utility evaluation typically involves training models on synthetic data and evaluating their performance on a holdout set of real data~\cite{ctabgan}, to enable sharing data while preserving privacy, for example with the goal of spurring research on a particular task described by a private and otherwise inaccessible dataset. 
In such cases, say when sharing synthetic data for competitions and shared tasks, it may be preferable to perform both training and evaluation on synthetic data, to allow participants to iterate on their solutions. Therefore, it is also important that the relative performance of different approaches trained and evaluated on synthetic data (train synthetic, test synthetic -- TSTS) is preserved when compared to real data performance evaluations (train synthetic, test real -- TSTR, or, train real, test real -- TRTR). For instance, if method A outperforms method B when trained and evaluated on real data, the same ranking should hold when both methods are trained and evaluated on synthetic data~\cite{Jordon2018PATE-GAN:Guarantees}.

Specific utility targets are typically dataset-dependant. For tabular data that serves a specific task (e.g., credit card fraud~\cite{Pozzolo2015CalibratingClassification}), it is reasonable to evaluate the performance of classifiers trained to predict values from the label column, but more generally, any column can be selected as classification/regression target, assuming a meaningful correlation with other features exists. For text and images, down-stream tasks almost exclusively concern classification accuracy~\cite{Yue2023SyntheticRecipe,Harder2021DP-MERF:Generation,Harder2023Pre-trainedGeneration}. Meanwhile some works have also proposed to evaluate the language modelling performance of models trained on down-stream synthetic data, by means of perplexity for next-token-prediction (decoder) models~\cite{Flemings2024DifferentiallyGeneration} and accuracy for fill-in-the-mask (encoder) models~\cite{Xie2024DifferentiallyText}.


\subsection{Privacy}
Privacy evaluation encompasses both formal guarantees and empirical privacy assessments through membership inference attacks (MIA)~\cite{Shokri2017MembershipModels}. Formal DP guarantees provide worst-case privacy bounds, ensuring a mathematically sound level of privacy protection. In contrast, empirical MIA evaluations measure the practical success rate of attempts to determine whether specific records were used to generate a synthetic data set. While formal guarantees provide absolute privacy assurance, MIA results demonstrate the current achievable privacy leakage under specific attack strategies. However, it is important to note that MIA evaluations do not guarantee protection against future, more sophisticated attack methods that might approach the theoretical bounds established by formal guarantees. A gold-standard approach to evaluate the susceptibility of a mechanism to MIA--- the ``shadow-modelling technique'' mentioned in Section~\ref{sec:background}---is to generate a dataset of triples consisting of a record, the mechanism's output and a label whether the record was used in producing the output or not~\cite{Carlini2022MembershipPrinciples}. A meta-classifier is then trained to predict the label, given record and output~\cite{stadler2022synthetic,Guepin2023SyntheticData}. 
Based on the assumptions about the attackers knowledge, additional auxilliary information can be provided, such as access to the generator model~\cite{Houssiau2022TAPAS:Data}. Since, in the context of differentially-private synthetic data generation, the output is a dataset, privacy evaluation at that level, especially for large unstructured datasets and models, is costly, as it would involve the generation of many synthetic datasets to train the meta-classifier. Instead, many works either rely only on formal guarantees~\cite{Jordon2018PATE-GAN:Guarantees,Lin2024DifferentiallyImages}\footnote{Which might not hold up in reality, for example due to false assumptions or implementation errors~\cite{Ganev2025TheDebugging}} or perform MIA with weak assumptions, e.g., by assuming that an attacker would only have access to models trained on synthetic data~\cite{Carlini2021ExtractingModels,Mattern2023MembershipComparison}, rather than to the synthetic data itself, as is the case with most recent works on DP text generation~\cite{Xie2024DifferentiallyText,Mattern2022DifferentiallySharing}.
\subsection{Benchmarks and Datasets}
\label{sec:datasets}
Across the different modalities, various datasets have been used as evaluation benchmarks for different proposed synthetic data generation approaches. It is worth pointing out, that a standardised benchmark suite is lacking.

For tabular data, various datasets have been used in different studies. Datasets are typically selected from purportedly critical domains, such as medical~\cite{Johnson2016MIMIC-IIIDatabase}, financial~\cite{Pozzolo2015CalibratingClassification}, as well as census data~\cite{Ruggles20156.0}. For images, synthetic data is derived from established resources, such as CIFAR~\cite{krizhevsky2009learning}, MNIST~\cite{lecun1998mnist}, FashionMnist~\cite{Xiao2017Fashion-MNIST:Algorithms}, CelebA~\cite{liu2015faceattributes} and other popular image datasets. The case is similar with synthetic text data, where popular sentiment analysis datasets, such as Yelp, IMDb or Amazon~\cite{Blitzer2007BiographiesClassification} reviews are used as ``private data'' to synthesise datasets from. For class-unconditioned datasets, the PubMed abstract database~\cite{canese2013pubmed} and various datasets usually used for sentiment analysis or topic modelling~\cite{Zhang2015Character-levelClassification} have been utilised.  







\section{Challenges and Future Research Avenues}
\label{sec:challenges}

After presenting a brief summary of the existing body of literature we present the current challenges in the field and problems that remain unaddressed, as a set of suggestions for potential future research avenues. More specifically, we relate our analysis to the current trends in the wider Machine Learning and Artificial Intelligence research areas---the focus on unstructured data and the shift to large-scale generative models pre-trained on large auxiliary corpora, anticipating that these developments will further advance the field of privacy-preserving synthetic data generation. 

\subsection{The need for Realistic Benchmarks}

\subsubsection{Scalability}
Despite significant advancements in tabular data synthesis, challenges to scalability persist. For single-table generation, many state-of-the-art (SOTA)~\cite{ctgan,ctabgan,zhao2023fct,tabddpm} models rely on mode-specific normalisation for columns containing non-categorical values, which  involves estimating a variational Gaussian mixture (VGM) model. However, this estimation process is computationally expensive for large datasets, thus limiting scalability. Furthermore, certain state-of-the-art models~\cite{zhang2024mixedtype,solatorio2023realtabformer,great,tabula,tabddpm} that use diffusion or LLMs, owing to their size, exhibit much longer training and generation times compared to GAN- or VAE-based methods that are typically much smaller and trained for specific tasks. While these models achieve better performance on small to medium datasets, their extended computational requirements can make them impractical for large-scale datasets that need to be generated on-demand.

In the context of relational table generation, i.e. generating multiple tabular datasets where entries (rows) are interlinked, addressing composite key constraints and sequential dependencies remains a persistent challenge~\cite{fakedb,fkgan,solatorio2023realtabformer}. Preliminary solutions~\cite{li2023irg} have been proposed to tackle these issues; however, they are not yet scalable to large relational databases, leaving room for further research and development in this area.

\subsubsection{Representativeness}
As outlined in Section~\ref{sec:datasets}, there is no consensus on what constitutes a good evaluation benchmark for synthetic data generation; as a result a variety of datasets is being used, exacerbating the comparison of proposed approaches. As such, the research field is missing out on potentially accelerated progress associated with a clear understanding of the limitations of the current state of the art and subsequent targeted improvements thereon, facilitated by the availability of easily accessible, high-quality standardised benchmarks and robust evaluation suites---the very reason we attribute to the rapid advancement of AI technology in the introduction of this paper.

Specifically for the unstructured modalities the current choice of evaluation datasets is additionally concerning, a fact that the community is beginning to realise~\cite{Tramer2024Position:Pretraining}. The chosen benchmarks represent general domains---tasks (such as sentiment analysis) associated with text and images widely available on the world wide web. Firstly, with synthetic data generation approaches increasingly relying on large pre-trained foundational models~\cite{Yue2023SyntheticRecipe,Mattern2022DifferentiallySharing,Harder2023Pre-trainedGeneration} which rely on ever-larger pre-trained corpora, the probability of encountering the benchmark data in the pre-training increases (data ``contamination''). As such, the allocated privacy budget might be compromised, as the exposure of the foundational model to the ``private'' data might not be accounted for. \citeauthor{Xie2024DifferentiallyText}~\cite{Xie2024DifferentiallyText} try to circumvent this fact by using an evaluation dataset that was gathered \emph{after} the foundational model underpinning their approach was used. However, this solution does not address the second, more fundamental issue: the focus on accessibility, reproducibility and ease of access biases the choice of benchmark data towards the public domain---and it is not necessarily the case that methods shown to work well on the general domain would generalise to specialised domains, each  with their specific requirements~\cite{DellAcqua2023NavigatingQuality,nagar2024llms}; this is even true within single domains, such as healthcare \cite{dizon2016adopt}, where local practice might depend on local demographics. Therefore, the areas of application (domains) that could benefit the most from privacy-preserving data sharing methods are excluded by design, due to sensitivity of the data in those domains 
For example, a realistic benchmark for generating synthetic hospital records might need to include actual hospital records as a ``private'' dataset, which is at odds with the stringent data sharing requirements associated with healthcare data as it contains potentially sensitive personally identifiable information.

To solve the former problem, we advocate to include experiments with \emph{fully open foundational models}~\cite{Groeneveld2024OLMo:Models-abbr} with their corresponding training corpora~\cite{Soldaini2024Dolma:Research-abbr} when developing methods that rely on foundational models, as it enables to verify if a benchmark's ``private'' data has been exposed to the model during pre-training. For the latter, we advocate to include resources with a ``gated'' resource mechanism, such as the requirement to sign a Data Usage Agreement (DUA), at the expense of the convenience of instantaneous access. Many datasets that might be prone to membership inference attacks and thus leak sensitive medical information have been made public under this access model~\cite{Johnson2016MIMIC-IIIDatabase,Johnson2023MIMIC-IVDataset,Zolnoori2019,grabar-etal-2018-cas,iso2017ntcir13} and their access has been streamlined by annotation standardisation efforts~\cite{Fries2022BigBio:Processing-abbr}. For researchers publishing new resources, we recommend to hold out a portion of the data from public release, which was a popular data release mechanism~\cite{rajpurkar2016squad,rajpurkar2018know,Yang2018,Dua2019} during the advent of pre-trained language models that required task-specific fine-tuning~\cite{Devlin2018}. While initially this practice was followed to ensure fair competition, during the testing of privacy-preserving synthetic data generation algorithms, it ensures that the data does not land in the public domain, pre-empting concerns on data contamination, as well as the sharing of sensitive contents. On the downside, such a sharing mechanism requires active maintenance.

\subsubsection{Multimodality}

The approaches covered in Section~\ref{sec:methods} are largely concerned with a distinct modality, i.e., tabular data, images or text in isolation. Meanwhile, in reality, data of different modalities might have intricate relationships, for example, textual radiology reports relate to images~\cite{Johnson2019MIMIC-CXRReports}. Similarly, many structured databases contain textual comments or links to images. There is no straightforward way to adopt any of the existing approaches to this multi-modal setting. One possible future research direction is to consider the multi-modal nature of some ``private'' datasets as a fundamental requirement. 



\subsection{Privacy}

\subsubsection{Unaccounted Privacy Leakage}
A common assumptions for generating class-conditioned data for down-stream classification tasks, is that the label distribution is publicly known~\cite{Mattern2022DifferentiallySharing,Yue2023SyntheticRecipe,Xie2024DifferentiallyText,Lin2024DifferentiallyImages}. However, this assumption is challenging when the number of labels increases---for example, 67\% of the records in the Mimic-III database~\cite{Johnson2016MIMIC-IIIDatabase} can be identified by a unique combination of a handful of the 50 most frequently occurring labels. Thus, an attacker would be able to mount a successful membership inference attack just considering labels alone. Indeed, it has been shown that these outliers with a combination of rare characteristics (e.g., rare diseases) are most susceptible to privacy leakage~\cite{Meeus2024AchillesPublishing}. To alleviate this issue, labels themselves should be generated via a differentially private mechanism, potentially jointly with the corresponding texts/images, which touches upon the challenge of multimodal data generation mentioned above.

More generally, many proposed approaches and algorithms rely on a set of hyper-parameters (e.g., sampling temperature for next-token prediction in text generation~\cite{Xie2024DifferentiallyText}), which depends on the task and model. Since they are therefore dependant on the private dataset, they should be estimated in a differentially private way~\cite{Gupta2010DifferentiallyOptimization,Abadi2016DeepPrivacy}, but very few papers\footnote{such as e.g.,~\citeauthor{Gondara2020DifferentiallyProjections}~\cite{Gondara2020DifferentiallyProjections} who discount 5\% of the privacy budget for hyper-parameter estimation} have concerned themselves with this fact.


\subsubsection{Empirical verification of formal guarantees}
Due to the fact that the gold standard approach of auditing DP guarantees with MIA essentially requires the generation of many datasets to train a meta-classifier, many works either omit a rigorous empirical evaluation of the privacy leakage of their proposed approaches~\cite{Zhang2017PrivNetworks,Jordon2018PATE-GAN:Guarantees,Lin2024DifferentiallyImages} or substitute it for simpler checks, such as evaluating the privacy leakage of down-stream models trained on the synthetic data~\cite{Xie2024DifferentiallyText} or doing anecdotal checks, such as inserting ``canary'' instances into the private data and then tracing their appearance in the synthesised datasets~\cite{Yue2023SyntheticRecipe}. We argue that conducting membership inference attacks is important for two reasons.

First, it validates that the implementation of the proposed algorithm is correct, which in fact does not always seem to be the case~\cite{Ganev2025TheDebugging}. One might argue that it is enough to prove the formal guarantees in theory, and the implementation is a separate matter. However, fidelity and utility measures are being established empirically, thus indeed relying on the implementation, which, if faulty, could invalidate the reported scores~\cite{Ganev2025TheDebugging}.

Second, while the guaranteed privacy budget $\epsilon$ corresponds to the worst-case MIA attack success rate, it represents a ``true positive rate at false positive rate'' curve~\cite{Carlini2021ExtractingModels}. In other words, while the precision and recall of an attack can be converted to a single $\epsilon$ value, the converse is not true, as a single epsilon value equals to any number on the curve (representing the worst case scenario)\footnote{following the equation $\epsilon = \max(\log{\frac{1-\delta-FPR}{FNR}, \frac{1-\delta-FNR}{FPR})}$ from~\citeauthor{Kairouz2015ThePrivacy}~\cite{Kairouz2015ThePrivacy}}. This is an important distinction, because an attack that has a 90\% re-identification accuracy on 1\% of the dataset arguably leads to worse consequences than an attack that has 50.1\% accuracy (slightly above random guess) on 50\% of the dataset. While both might be possible under a worst-case scenario, only an empirical evaluation based on a real MIA can show where the worst-case bound can be reached.

To address this shortcoming, we suggest to include diagnostic MIA attacks when evaluating DP data generation algorithms in form of DP audits~\cite{Ganev2025TheDebugging}. To alleviate the computational cost, these can be mounted on toy data to validate that the implementation is sound in principle.



\section{Empirical validation of the challenges}
\label{sec:validation}
To validate (some of) the technical challenges identified in the previous section, we conduct an empirical study, focussing on emerging generative approaches. Specifically, we benchmark four state-of-the-art methods that rely on transfer learning from web-scale corpora in order to synthesise unstructured data while providing differential privacy guarantees. We focus on \emph{realistic} privacy budgets ($\epsilon \in \{0.5, 1, 2, 4\}$) and benchmark on challenging datasets with gated access, which makes it less likely that they appear in the pre-training data of large generative foundational models

\subsection{Experiment Setup}

\begin{wrapfigure}{R}{0.6\textwidth}
\centering
\vspace{-1.5em}
\captionof{table}{F1 scores for different approaches, down-stream models, and privacy budgets ($\epsilon$) across text datasets.}
\label{tab:text-utility}
\resizebox{0.6\textwidth}{!}{

\begin{tabular}{llccccc}
\toprule
\makecell{Downsteam\\Model} & Method & $\epsilon = \infty$ & $\epsilon = 4$ & $\epsilon = 2$ & $\epsilon = 1$ & $\epsilon = 0.5$ \\
\midrule
\rowcolor{black!10}
\multicolumn{7}{c}{\textsc{Hallmarks Of Cancer}} \\
\multirow{3}{*}{\makecell{\textsc{BERT-}\\\textsc{large}}} & Original & 71.9\\
 & DP-Gen & 54.8 & 19.2 & 14.7 & 13.6 & 17.1 \\
 & AUG-PE & 15.0 & 8.2 & 6.6 & 7.6 & 5.0 \\
\multirow{3}{*}{\makecell{\textsc{Bio-}\\\textsc{ClinicalBERT}}} & Original & 68.7 \\
 & DP-Gen & 48.9 & 15.8 & 11.7 & 9.2 & 4.5 \\
 & AUG-PE & 12.9 & 6.8 & 7.4 & 8.3 & 6.5 \\
\multirow{3}{*}{\makecell{\textsc{DeBERTa-}\\\textsc{xlarge}}} & Original & 52.2\\
 & DP-Gen & 39.0 & 0.0 & 0.0 & 0.0 & 20.7 \\
 & AUG-PE & 2.8 & 0.4 & 11.9 & 11.3 & 3.8 \\
\midrule
\rowcolor{black!10}
\multicolumn{7}{c}{\textsc{N2C2 2008}} \\
\multirow{3}{*}{\makecell{\textsc{LongFormer-}\\\textsc{large}}} & Original & 87.7  \\
 & DP-Gen & 61.1 & 55.8 & 59.3 & 56.9 & 55.6 \\
 & AUG-PE & 58.1 & 58.9 & 53.2 & 59.1 & 55.9 \\
\multirow{3}{*}{\makecell{\textsc{Clinical-}\\\textsc{BigBird}}} & Original & 71.6 \\
 & DP-Gen & 55.7 & 53.2 & 53.2 & 53.2 & 53.4 \\
 & AUG-PE & 53.2 & 53.2 & 53.2 & 53.2 & 53.2 \\
\multirow{3}{*}{\makecell{\textsc{Clinical-}\\\textsc{LongFormer}}} & Original & 60.1 \\
 & DP-Gen & 59.0 & 55.4 & 53.6 & 53.2 & 53.2 \\
 & AUG-PE & 53.2 & 53.2 & 53.2 & 53.2 & 56.9 \\
\midrule
\rowcolor{black!10}
\multicolumn{7}{c}{\textsc{PsyTAR}} \\
\multirow{3}{*}{\makecell{\textsc{BERT-}\\\textsc{base}}} & Original & 79.7 \\
 & DP-Gen & 69.5 & 33.5 & 32.2 & 33.3 & 31.1 \\
 & AUG-PE & 61.0 & 62.1 & 60.9 & 54.5 & 49.5 \\
\multirow{3}{*}{\makecell{\textsc{BERT-}\\\textsc{large}}} & Original & 80.4\\
 & DP-Gen & 70.3 & 39.1 & 36.0 & 36.1 & 31.7 \\
 & AUG-PE & 63.9 & 64.7 & 63.5 & 58.1 & 50.9 \\
\multirow{3}{*}{\makecell{\textsc{DeBERTa-}\\\textsc{xlarge}}} & Original & 82.1\\
 & DP-Gen & 75.3 & 23.9 & 15.8 & 27.6 & 4.9 \\
 & AUG-PE & 44.0 & 65.3 & 65.7 & 60.2 & 54.4 \\
\midrule
\bottomrule
\end{tabular}

}
\vspace{-2em}
\end{wrapfigure}
\sloppy

\paragraph{Benchmarked Approaches:} We benchmark the following four approaches. For detailed descriptions of our adaptations, consult the Appendix. If not otherwise specified, we use \texttt{Meta-LLama-3.2-1B} as the generative model for text and \texttt{StableDiffusion-2.5-XL} for images.

\textbf{AUG-PE}~\cite{Xie2024DifferentiallyText}
(Augmented Private Evolution) generates differentially private synthetic text without model training, requiring only API access to large language models. It iteratively selects synthesized texts most similar to private data and creates variations through prompting techniques, such as paraphrasing. Privacy is ensured by adding Gaussian noise to the comparison process.

\textbf{DP Finetune}~\cite{Yue2023SyntheticRecipe}: This approach fine-tunes pre-trained generative language models (e.g., GPT-2) with Differentially Private Stochastic Gradient Descent (DP-SGD) on private data. The resulting DP-trained model is then used to generate synthetic text.

\textbf{DPSDA}~\cite{Lin2024DifferentiallyImages} (Differentially Private Synthetic Data via APIs)
implements the Private Evolution (PE) for images using blackbox API access to foundation models like diffusion models. This is the original version of the AUG-PE algorithm, initially applied to generate private images.

\textbf{DP-MERF}~\cite{Harder2021DP-MERF:Generation} DP-MEPF (Differentially Private Mean Embeddings with Perceptual Features) generates synthetic images by leveraging perceptual features from models pre-trained on public data. It constructs kernel mean embeddings using these features, privatizes them once with the Gaussian mechanism, and then trains a generator to minimize the maximum mean discrepancy between synthetic and private distributions. Unlike adversarial approaches, this non-adversarial method allows for stable optimization and does not accumulate privacy loss during training, as the privatized mean embedding is used repeatedly. We don't use a diffusion model here, in line with the original implementation.

\paragraph{``Private'' Datasets:} We choose datasets from the medical domain that are unlikely to be found in the general domain. For text, we use the \textsc{Hallmarks of Cancer}~\cite{DBLP:journals/bioinformatics/BakerSGAHSK16}, \textsc{n2c2 2008}~\cite{uzuner2009recognizing} and \textsc{PsyTAR}~\cite{Zolnoori2019} datasets. \textsc{HoC}, a multilabel classification dataset with heavy imbalance towards the ``no mention'' class, is publicly available (thus potentially observed during pre-training) but has the challenge of domain specificity, as it requires to identify mentions of specific cancer hallmarks in sentences from scientific paper abstracts. \textsc{n2c2 2008} features the multi-label task of recognising obesity and various co-morbidities from long-context MIMIC-III discharge summaries. \textsc{PsyTAR} requires to detect mentions of various adverse drug effects from social media posts as a multi-label task. Both \textsc{n2c2 2008} and \textsc{PsyTAR} are only accessible via a gated mechanism, thus unlikely appearing ``in the wild'' in a web scrape. For images, we include the \textsc{FracAtlas}~\cite{Abedeen2023FracAtlas:Radiographs} and \textsc{SiPaKMeD}~\cite{Plissiti2018Sipakmed:Images} datasets. \textsc{FracAtlas} is a collection of X-Ray images with various annotations. We use the fractured and non-fractured annotations to construct a (heavily imbalanced) binary classification task. \textsc{SiPaKMeD} is a dataset of pap smear blood film images with blood cell annotations. We use the cropped images that feature a single blood cell per image and use the five different blood cell types as labels.

\paragraph{Evaluation Protocol:} Unless otherwise indicated, we generate synthetic data from the training portion of the selected datasets using the selected approaches. In line with existing literature, we assume label distributions to be public and condition the generation upon them. For multi-label data we treat each unique combination of labels as a separate label. We keep the class composition of synthetic data equal to that of the original training set (bar some rounding errors due to batching).

As a measure of fidelity, we evaluate the FID score for generated images and the MAUVE score and divergence\footnote{we measure KL divergence unless stated otherwise} of the distribution of recognised named entities. We further report various statistical measures---channel-wise distribution differences for images and text length and collocation divergences for text. Specifically, we evaluate the pairwise scores or divergences between synthetic train and original test data, and compare it to scores between original train and test.
For utility, we evaluate the down-stream performance of different classification models trained on synthetic data and evaluated on real data. All measures are evaluated at $\epsilon \in \{\infty, 0.5, 1,2,4\}$.


\subsection{Results \& Analysis}
We present the utility and fidelity of the generated data and showcase randomly chosen generations for a qualitative impression. 

\begin{table}[t]
\centering
\caption{Performance metrics for different approaches, down-stream models, and privacy budgets ($\epsilon$) across image datasets. Each row shows Accuracy and Matthews Correlation Coefficient (MCC) scores.}
\label{tab:image-utility}
\resizebox{.85\textwidth}{!}{
\begin{tabular}{lllrlrlrlrlr}
\toprule
\multirow{2}{*}{\makecell{Downsteam\\ Model}} & \multirow{2}{*}{Method} & \multicolumn{2}{c}{$\epsilon = \infty$} & \multicolumn{2}{c}{$\epsilon = 4$} & \multicolumn{2}{c}{$\epsilon = 2$} & \multicolumn{2}{c}{$\epsilon = 1$} & \multicolumn{2}{c}{$\epsilon = 0.5$} \\
 &  & \multicolumn{2}{c}{\scorepair{Acc}{MCC}}   & \multicolumn{2}{c}{\scorepair{Acc}{MCC}}  & \multicolumn{2}{c}{\scorepair{Acc}{MCC}}  & \multicolumn{2}{c}{\scorepair{Acc}{MCC}}  & \multicolumn{2}{c}{\scorepair{Acc}{MCC}}  \\
\midrule
\rowcolor{black!10}
\multicolumn{12}{c}{\textsc{FracAtlas}} \\
\multirow{3}{*}{\textsc{ViT-Base}} & \multirow{1}{*}{Original} & \multicolumn{2}{c}{\scorepair{$92.1$}{$71.8$}} &  &  &  &  \\
  & \multirow{1}{*}{DP-MEPF} & \multicolumn{2}{c}{\scorepair{$86.4$}{$48.5$}} & \multicolumn{2}{c}{\scorepair{$82.4$}{$0.0$}} & \multicolumn{2}{c}{\scorepair{$82.4$}{$0.0$}} & \multicolumn{2}{c}{\scorepair{$82.4$}{$0.0$}} & \multicolumn{2}{c}{\scorepair{$47.6$}{$6.6$}} \\
  & \multirow{1}{*}{DPSDA} & \multicolumn{2}{c}{\scorepair{$62.5$}{$10.7$}} & \multicolumn{2}{c}{\scorepair{$82.4$}{$0.0$}} & \multicolumn{2}{c}{\scorepair{$37.0$}{$4.3$}} & \multicolumn{2}{c}{\scorepair{$72.1$}{$21.2$}} & \multicolumn{2}{c}{\scorepair{$82.5$}{$7.8$}} \\
\multirow{3}{*}{\textsc{ViT-Large}} & \multirow{1}{*}{Original} & \multicolumn{2}{c}{\scorepair{$92.5$}{$72.7$}} &  &  &  &  \\
  & \multirow{1}{*}{DP-MEPF} & \multicolumn{2}{c}{\scorepair{$85.3$}{$50.8$}} & \multicolumn{2}{c}{\scorepair{$82.4$}{$0.0$}} & \multicolumn{2}{c}{\scorepair{$82.4$}{$0.0$}} & \multicolumn{2}{c}{\scorepair{$82.4$}{$0.0$}} & \multicolumn{2}{c}{\scorepair{$81.4$}{$-2.2$}} \\
  & \multirow{1}{*}{DPSDA} & \multicolumn{2}{c}{\scorepair{$32.6$}{$9.8$}} & \multicolumn{2}{c}{\scorepair{$56.4$}{$4.3$}} & \multicolumn{2}{c}{\scorepair{$71.3$}{$3.3$}} & \multicolumn{2}{c}{\scorepair{$71.4$}{$8.9$}} & \multicolumn{2}{c}{\scorepair{$77.1$}{$1.7$}} \\
\multirow{3}{*}{\textsc{BEiT-Base}} & \multirow{1}{*}{Original} & \multicolumn{2}{c}{\scorepair{$92.8$}{$73.5$}} &  &  &  &  \\
  & \multirow{1}{*}{DP-MEPF} & \multicolumn{2}{c}{\scorepair{$84.5$}{$50.7$}} & \multicolumn{2}{c}{\scorepair{$82.4$}{$0.0$}} & \multicolumn{2}{c}{\scorepair{$82.4$}{$0.0$}} & \multicolumn{2}{c}{\scorepair{$82.4$}{$0.0$}} & \multicolumn{2}{c}{\scorepair{$82.4$}{$0.0$}} \\
  & \multirow{1}{*}{DPSDA} & \multicolumn{2}{c}{\scorepair{$69.2$}{$16.3$}} & \multicolumn{2}{c}{\scorepair{$76.0$}{$3.9$}} & \multicolumn{2}{c}{\scorepair{$82.8$}{$17.9$}} & \multicolumn{2}{c}{\scorepair{$79.0$}{$12.3$}} & \multicolumn{2}{c}{\scorepair{$81.2$}{$0.4$}} \\
\midrule
\rowcolor{black!10}
\multicolumn{12}{c}{\textsc{SIPaKMeD}} \\
\multirow{3}{*}{\textsc{ViT-Base}} & \multirow{1}{*}{Original} & \multicolumn{2}{c}{\scorepair{$97.8$}{$97.2$}} &  &  &  &  \\
  & \multirow{1}{*}{DP-MEPF} & \multicolumn{2}{c}{\scorepair{$77.1$}{$71.7$}} & \multicolumn{2}{c}{\scorepair{$43.6$}{$37.5$}} & \multicolumn{2}{c}{\scorepair{$41.9$}{$28.6$}} & \multicolumn{2}{c}{\scorepair{$30.4$}{$13.8$}} & \multicolumn{2}{c}{\scorepair{$28.0$}{$10.7$}} \\
  & \multirow{1}{*}{DPSDA} & \multicolumn{2}{c}{\scorepair{$28.0$}{$10.3$}} & \multicolumn{2}{c}{\scorepair{$19.5$}{$-1.2$}} & \multicolumn{2}{c}{\scorepair{$23.5$}{$4.3$}} & \multicolumn{2}{c}{\scorepair{$27.6$}{$15.4$}} & \multicolumn{2}{c}{\scorepair{$20.9$}{$1.0$}} \\
\multirow{3}{*}{\textsc{ViT-Large}} & \multirow{1}{*}{Original} & \multicolumn{2}{c}{\scorepair{$98.5$}{$98.2$}} &  &  &  &  \\
  & \multirow{1}{*}{DP-MEPF} & \multicolumn{2}{c}{\scorepair{$81.8$}{$78.1$}} & \multicolumn{2}{c}{\scorepair{$57.0$}{$48.4$}} & \multicolumn{2}{c}{\scorepair{$50.1$}{$39.7$}} & \multicolumn{2}{c}{\scorepair{$46.1$}{$36.0$}} & \multicolumn{2}{c}{\scorepair{$29.4$}{$14.4$}} \\
  & \multirow{1}{*}{DPSDA} & \multicolumn{2}{c}{\scorepair{$25.2$}{$8.8$}} & \multicolumn{2}{c}{\scorepair{$26.2$}{$9.2$}} & \multicolumn{2}{c}{\scorepair{$25.7$}{$8.2$}} & \multicolumn{2}{c}{\scorepair{$23.0$}{$4.5$}} & \multicolumn{2}{c}{\scorepair{$20.0$}{$-0.5$}} \\
\multirow{3}{*}{\textsc{BEiT-Base}} & \multirow{1}{*}{Original} & \multicolumn{2}{c}{\scorepair{$96.2$}{$95.2$}} &  &  &  &  \\
  & \multirow{1}{*}{DP-MEPF} & \multicolumn{2}{c}{\scorepair{$82.8$}{$78.9$}} & \multicolumn{2}{c}{\scorepair{$56.7$}{$48.2$}} & \multicolumn{2}{c}{\scorepair{$63.1$}{$55.4$}} & \multicolumn{2}{c}{\scorepair{$44.1$}{$36.5$}} & \multicolumn{2}{c}{\scorepair{$24.0$}{$6.2$}} \\
  & \multirow{1}{*}{DPSDA} & \multicolumn{2}{c}{\scorepair{$30.8$}{$15.6$}} & \multicolumn{2}{c}{\scorepair{$28.4$}{$11.2$}} & \multicolumn{2}{c}{\scorepair{$27.2$}{$9.9$}} & \multicolumn{2}{c}{\scorepair{$23.2$}{$4.0$}} & \multicolumn{2}{c}{\scorepair{$21.3$}{$1.6$}} \\

\midrule
\bottomrule
\end{tabular}

}
\end{table}

\paragraph{Utility:} Tables~\ref{tab:text-utility}~and~\ref{tab:image-utility} present the utility of the down-stream data. It becomes apparent that the performance of the selected approaches deteriorates significantly compared to the evaluations selected in the respective studies. For context, text methods achieve only $84\%, 46\%, 34\%, 43\%, 43\%$ (DP-Generator) and $56\%, 57\%, 58\%, 56\% 51\%$ (AUG-PE) at $\epsilon \in {\infty, 0.5, 1, 2, 4}$ of the TRTR model performance on average, compared to almost outperforming the baseline as reported on the simpler public domain datasets~\cite{Yue2023SyntheticRecipe,Xie2024DifferentiallyText}. Strikingly, this deterioration already becomes apparent at $\epsilon = \infty$ without any privacy guarantees, suggesting that the underlying language model indeed struggles to capture the intricacies of the domain.

\begin{figure}[tbh]

\centering

    \resizebox{.65\textwidth}{!}{\input{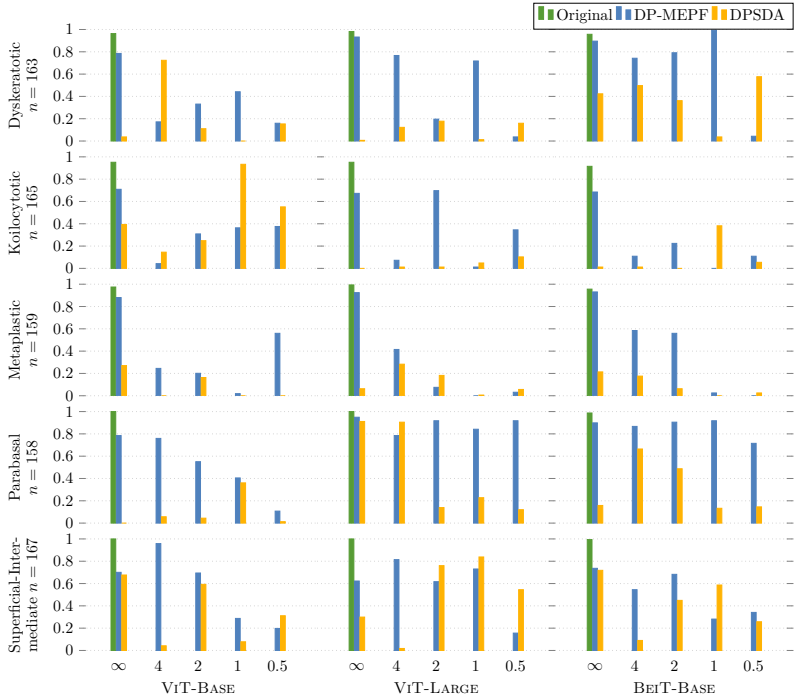}}
    \caption{Per-class accuracies of all models trained on original and synthetic \textsc{SIPaKMeD} data, generated with varying $\epsilon$ budgets.}
    \label{fig:img-per-cls-sipakmed}
\end{figure}

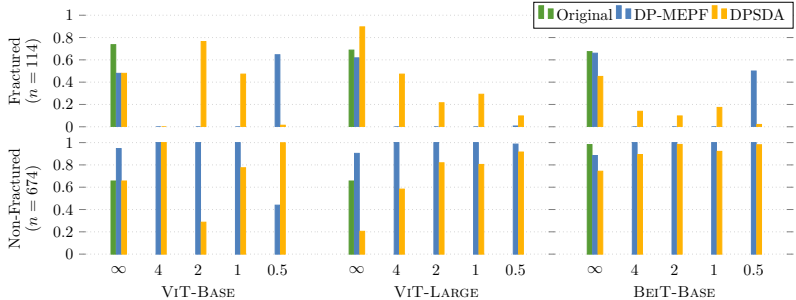
\begin{figure}[tbh]

\centering

    \resizebox{.65\textwidth}{!}{\begin{tikzpicture}
\begin{groupplot}[
       group style={
       group name=plot,
       group size=3 by 2,
       xlabels at=edge bottom,
       ylabels at=edge left,
       horizontal sep=0pt,
       vertical sep=10pt,
       /pgf/bar width=3pt
       },
    major x tick style=transparent,
    ybar= \pgflinewidth,
    ymax=1,
    ymin=0,
    x axis line style={opacity=0},
    x tick label style={rotate=0, anchor=center},
    xticklabel style={yshift=-2mm,xshift={ifthenelse(\ticknum==2,0,0)}}, 
    xtick={1, 2, 3, 4, 5, 6},
    xticklabels={\empty, $\infty$, $4$, $2$, $1$, $0.5$},
    ymajorgrids=true,
    grid style=dotted,
    nodes near coords,
    scale only axis,
    point meta=explicit symbolic,
    enlarge x limits = {abs=1},
    legend cell align={left},
    legend style={font=},
    legend columns=-1,
    legend style={at={(1,1)},anchor=east},
    cycle list={
      draw=igreen,fill=igreen,nodes near coords style={igreen}\\
      draw=bblue,fill=bblue,nodes near coords style={bblue!60}\\
      draw=rred,fill=rred,nodes near coords style={rred}\\
      },
    legend columns=-1,
    height=0.15\textwidth,
    ]
    \nextgroupplot[
         width=0.32\textwidth,
         ytick pos=left,
         xtick=\empty,
         ylabel={\makecell{Fractured\\($n=114$)}}
          ]
    \addplot  coordinates {
      (2,0.7361111044883728)
      };
    \addplot  coordinates {
      (2,0.4791666567325592)
      (3,0)
      (4,0)
      (5,0)
      (6,0.6458333134651184)
    };
    \addplot  coordinates {
      (2,0.4791666567325592)
      (3,0)
      (4,0.7638888955116272)
      (5,0.4722222089767456)
      (6,0.013888888992369175)
    };
    
    \nextgroupplot[
         yticklabels={},
         width=0.32\textwidth,
         ytick pos=right,
         xtick=\empty
         ]
    \addplot coordinates {
      (2,0.6875)
      };
  \addplot  coordinates {
      (2,0.6180555820465088)
      (3,0)
      (4,0)
      (5,0)
      (6, 0.0069444444961845875)
    };
  \addplot  coordinates {
      (2,0.8958333134651184)
      (3,0.4722222089767456)
      (4,0.2152777761220932)
      (5,0.2916666567325592)
      (6,0.0972222238779068)
    };
    \nextgroupplot[
         yticklabels={},
         width=0.32\textwidth,
         ytick pos=right,
         xtick=\empty
         ]
    \addplot coordinates {
      (2,0.6736111044883728)
      };
  \addplot  coordinates {
      (2,0.6597222089767456)
      (3,0)
      (4,0)
      (5,0)
      (6, 0.5)
    };
  \addplot  coordinates {
      (2,0.4513888955116272)
      (3,0.1388888955116272)
      (4,0.0972222238779068)
      (5,0.1736111044883728)
      (6,0.02083333395421505)
    };

    \legend{Original, DP-MEPF, DPSDA}
    \nextgroupplot[
         xlabel=\textsc{ViT-Base},
         width=0.32\textwidth,
         ytick pos=left,
         ylabel={\makecell{Non-Fractured\\($n=674$)}}
          ]
    \addplot coordinates {
      (2,0.6557863354682922)
      };
    \addplot  coordinates {
      (2,0.9465875625610352)
      (3,1)
      (4,1)
      (5,1)
      (6, 0.43916913866996765)
      };
      
    \addplot  coordinates {
      (2,0.6557863354682922)
      (3,1)
      (4,0.28635016083717346)
      (5,0.7744807004928589)
      (6,0.998516321182251)
      };

    \nextgroupplot[
         xlabel=\textsc{ViT-Large},
         yticklabels={},
         width=0.32\textwidth,
         ytick pos=right
         ]
    \addplot coordinates {
     (2,0.6557863354682922)
      };
    \addplot  coordinates {
      (2,0.9035608172416687)
      (3,1)
      (4,1)
      (5,1)
      (6,0.9866468906402588)
      };
      
    \addplot  coordinates {
      (2,0.2047477811574936)
      (3,0.5830860733985901)
      (4,0.8189911246299744)
      (5,0.8041542768478394)
      (6,0.9154302477836609)
      };

  \nextgroupplot[
         xlabel=\textsc{BeiT-Base},
         yticklabels={},
         width=0.32\textwidth,
         ytick pos=right
         ]
    \addplot coordinates {
      (2,0.9821958541870117)
    };
    
    \addplot coordinates {
      (2,0.8842729926109314)
      (3,1)
      (4,1)
      (5,1)
      (6,1)
    };

    \addplot coordinates {
      (2,0.7433234453201294)
      (3,0.8931750655174255)
      (4,0.9836795330047607)
      (5,0.921364963054657)
      (6,0.9807121753692627)
    };

    \end{groupplot}
    
    \end{tikzpicture}}
    \caption{Per-class accuracies of all models trained on original and synthetic \textsc{FracAtlas} data, generated with varying $\epsilon$ budgets.}
    \label{fig:img-per-cls-fracatlas}
\end{figure}

\begin{table}[th]
    \centering
    \begin{minipage}{\textwidth}
    \centering
    \caption{Fidelity metrics for different approaches and privacy budgets ($\epsilon$) across text datasets. Each cell shows the MAUVE score and recognised Named Entity count KL-divergences.}
    \label{tab:text-fidelity}
        \resizebox{.85\textwidth}{!}{

\begin{tabular}{l c c c c c}
\toprule
\multirow{2}{*}{Method} & $\epsilon = \infty$ & $\epsilon = 4$ & $\epsilon = 2$ & $\epsilon = 1$ & $\epsilon = 0.5$ \\
 & \scorepair{MAUVE$\uparrow$}{NER$\downarrow$} & \scorepair{MAUVE$\uparrow$}{NER$\downarrow$} & \scorepair{MAUVE$\uparrow$}{NER$\downarrow$} & \scorepair{MAUVE$\uparrow$}{NER$\downarrow$} & \scorepair{MAUVE$\uparrow$}{NER$\downarrow$} \\
\midrule
\rowcolor{black!10}
\multicolumn{6}{c}{\textsc{Hallmarks Of Cancer}} \\
Original & \scorepair{$0.994$}{$1.043$} & & & & \\
DP-Gen & \scorepair{$0.012$}{$1.596$} & \scorepair{$0.011$}{$2.255$} & \scorepair{$0.011$}{$2.303$} & \scorepair{$0.011$}{$2.491$} & \scorepair{$0.011$}{$2.598$} \\
AUG-PE & \scorepair{$0.011$}{$2.740$} & \scorepair{$0.011$}{$3.279$} & \scorepair{$0.011$}{$3.443$} & \scorepair{$0.010$}{$4.188$} & \scorepair{$0.011$}{$4.831$} \\
\midrule
\rowcolor{black!10}
\multicolumn{6}{c}{\textsc{N2C2 2008}} \\
Original & \scorepair{$0.996$}{$0.739$} & & & & \\
DP-Gen & \scorepair{$0.135$}{$1.731$} & \scorepair{$0.032$}{$8.180$} & \scorepair{$0.034$}{$8.185$} & \scorepair{$0.023$}{$7.701$} & \scorepair{$0.033$}{$7.654$} \\
AUG-PE & \scorepair{$0.032$}{$8.700$} & \scorepair{$0.017$}{$8.849$} & \scorepair{$0.019$}{$8.735$} & \scorepair{$0.019$}{$8.937$} & \scorepair{$0.017$}{$8.858$} \\
\midrule
\rowcolor{black!10}
\multicolumn{6}{c}{\textsc{PsyTAR}} \\
Original & \scorepair{$0.988$}{$0.857$} & & & & \\
DP-Gen & \scorepair{$0.016$}{$1.846$} & \scorepair{$0.025$}{$2.025$} & \scorepair{$0.023$}{$2.115$} & \scorepair{$0.021$}{$2.229$} & \scorepair{$0.018$}{$2.401$} \\
AUG-PE & \scorepair{$0.019$}{$4.382$} & \scorepair{$0.020$}{$4.808$} & \scorepair{$0.019$}{$5.060$} & \scorepair{$0.019$}{$5.339$} & \scorepair{$0.017$}{$5.397$} \\
\bottomrule
\end{tabular}}
    \end{minipage}

    \begin{minipage}{\textwidth}
    \centering
    \caption{Fidelity metrics for different approaches and privacy budgets ($\epsilon$) across text datasets. Each cell shows n-gram distribution KL-divergences.}
    \label{tab:text-fidelity-2}
        \resizebox{.85\textwidth}{!}{\begin{tabular}{l c c c c c}
\toprule
\multirow{2}{*}{Method} & $\epsilon = \infty$ & $\epsilon = 4$ & $\epsilon = 2$ & $\epsilon = 1$ & $\epsilon = 0.5$ \\
 & n-gram 1|2|3$\downarrow$ & n-gram 1|2|3$\downarrow$ & n-gram 1|2|3$\downarrow$ & n-gram 1|2|3$\downarrow$ & n-gram 1|2|3$\downarrow$ \\
\midrule
\rowcolor{black!10}
\multicolumn{6}{c}{\textsc{Hallmarks Of Cancer}} \\
Original & \scoretriple{0.63}{2.50}{3.84} & & & & \\
DP-Gen & \scoretriple{1.55}{4.13}{5.83} & \scoretriple{1.79}{4.38}{5.57} & \scoretriple{1.88}{4.49}{5.50} & \scoretriple{1.91}{4.49}{5.67} & \scoretriple{1.99}{4.67}{5.65} \\
AUG-PE & \scoretriple{3.26}{5.90}{7.21} & \scoretriple{3.73}{6.20}{7.57} & \scoretriple{3.90}{6.35}{7.63} & \scoretriple{4.22}{6.68}{7.80} & \scoretriple{4.58}{7.32}{8.35} \\
\midrule
\rowcolor{black!10}
\multicolumn{6}{c}{\textsc{N2C2 2008}} \\
Original & \scoretriple{0.59}{1.46}{1.60} & & & & \\
DP-Gen & \scoretriple{1.39}{2.43}{2.81} & \scoretriple{7.23}{9.70}{9.02} & \scoretriple{7.40}{9.52}{8.88} & \scoretriple{6.92}{8.91}{7.77} & \scoretriple{6.81}{8.55}{7.66} \\
AUG-PE & \scoretriple{6.00}{9.47}{9.37} & \scoretriple{6.86}{9.78}{10.00} & \scoretriple{7.05}{9.69}{9.04} & \scoretriple{7.64}{10.33}{10.10} & \scoretriple{8.15}{10.31}{9.82} \\
\midrule
\rowcolor{black!10}
\multicolumn{6}{c}{\textsc{PsyTAR}} \\
Original & \scoretriple{0.61}{3.10}{6.32} & & & & \\
DP-Gen & \scoretriple{1.23}{5.87}{8.59} & \scoretriple{1.56}{5.40}{8.31} & \scoretriple{1.65}{5.41}{8.27} & \scoretriple{1.75}{5.58}{8.15} & \scoretriple{1.88}{5.75}{8.55} \\
AUG-PE & \scoretriple{3.56}{8.44}{11.58} & \scoretriple{3.58}{8.73}{11.53} & \scoretriple{4.00}{8.79}{11.65} & \scoretriple{4.37}{9.11}{11.58} & \scoretriple{4.54}{9.19}{11.62} \\
\bottomrule
\end{tabular}}
    \end{minipage}
    \vspace{-1em}
\end{table}

\begin{figure}[!thbp]
    \centering
    \includegraphics[width=\textwidth, clip, trim=0cm 0cm 0cm 4cm]{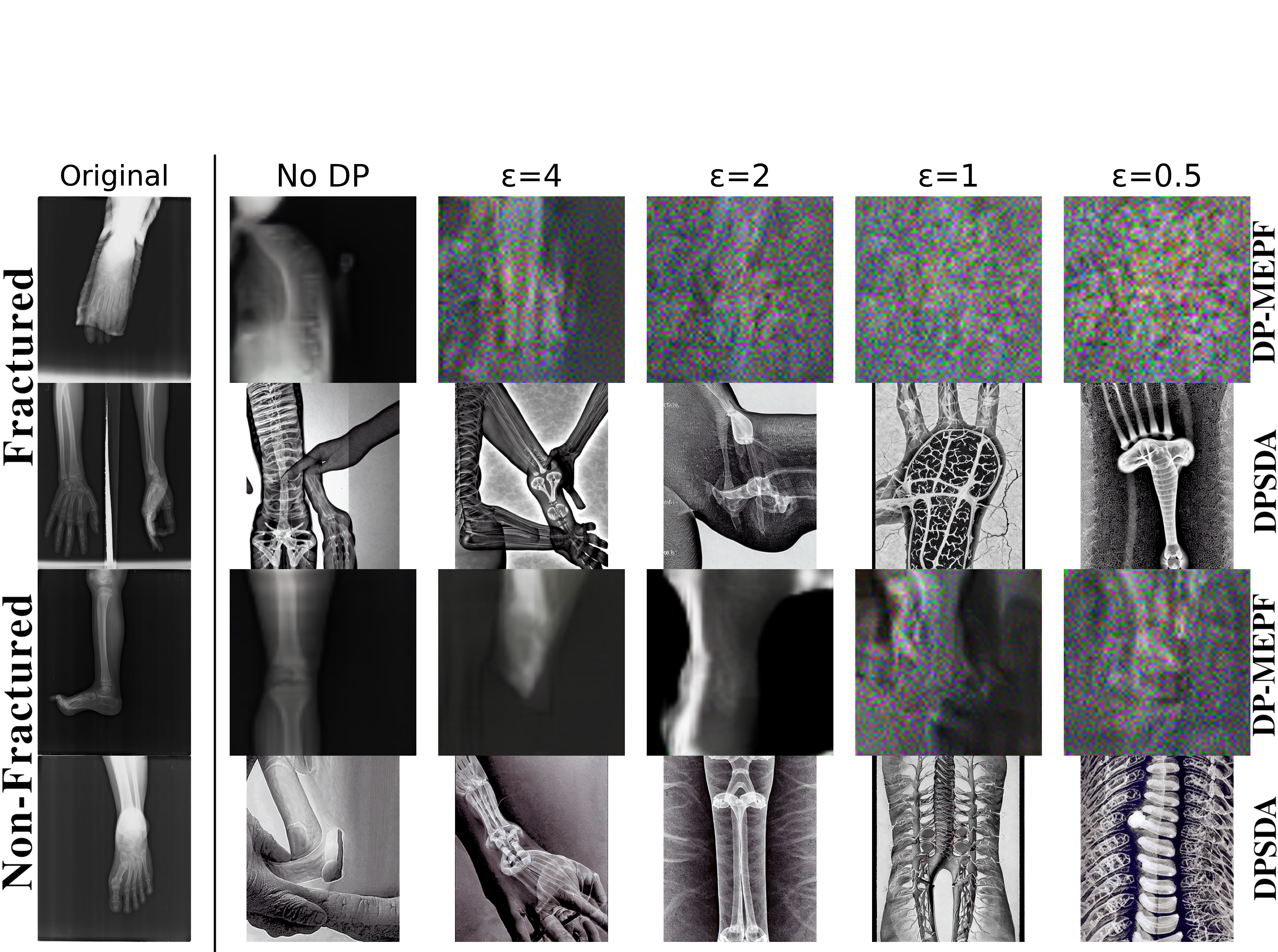}
    \caption{Random sample of \textsc{FracAtlas} images generated at different $\epsilon$. Note the manifestation of noise in DP-MEPF images with privacy guarantees and the low semantic coherence of DPSDA generations.}
    \label{fig:img-example}
\end{figure}

Similarly, the utility of image data also deteriorates quickly, leading to majority-class collapse of down-stream models on the unbalanced \textsc{FracAtlas} dataset (see e.g. Figure~\ref{fig:img-per-cls-fracatlas}, performance on the Fractured class). But even for the balanced \textsc{SIPaKMeD} dataset, the performance on some classes (see e.g., Figure~\ref{fig:img-per-cls-sipakmed} down-stream recognition performance of the the Metaplastic and Koilocytotic classes). Here, DP-MEPF is capable to capture the underlying distribution of private data much better when trained without added DP noise, but the addition of DP noise significantly impacts image quality, especially for the minority class. This is reasonable, as the noise is added directly to the data-dependant term, towards which the output distribution of the generator is optimised---thus, less data requires the addition of more noise to obfuscate each samples' contribution to the distribution (compare rows one and three in Figure~\ref{fig:img-example}).

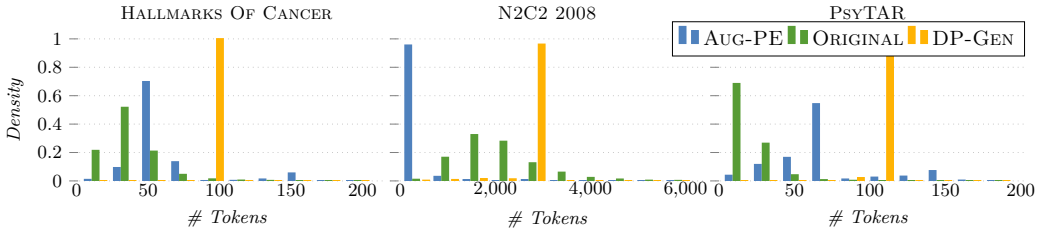
\begin{figure}[ht]
    \centering    
    \resizebox{.85\linewidth}{!}{%
        \begin{tikzpicture}[font=\footnotesize]
    \begin{groupplot}[
        group style={
            group size=3 by 1,
            horizontal sep=0.1cm,
            ylabels at=edge left,
            yticklabels at=edge left,
           /pgf/bar width=3pt
        },
        major x tick style=transparent,
        ybar= \pgflinewidth,
        ymax=1,
        ymin=0,
        x axis line style={opacity=0},
        x tick label style={rotate=0, anchor=center},
        ymajorgrids=true,
        grid style=dotted,
        width=0.42\textwidth,
        height=4cm,
        legend cell align={left},
        legend style={font=},
        legend columns=-1,
        legend style={at={(1,1)},anchor=east},
        ylabel=\emph{Density},
        xlabel=\emph{\# Tokens},
        no markers,
        xtick pos=left,
        ytick pos=left,
        every axis plot/.append style={thick},
        x axis line style={opacity=0},
    ]
    
    \nextgroupplot[title=\textsc{Hallmarks Of Cancer}]
    \addplot [mark=none, color=bblue, fill=bblue, legend entry=\textsc{aug-pe}] coordinates {
        (13.2, 0.008783201405152223) (33.599999999999994, 0.09221411475409837) (53.99999999999999, 0.6976981116315379) (74.39999999999999, 0.13319772131147542) (94.8, 0.0006840601092896175) (115.19999999999999, 0.0018550202966432475) (135.6, 0.011515441842310695) (156.0, 0.05406032864949258) (176.39999999999998, 1e-06) (196.8, 1e-06)
    };
    \addplot [mark=none, color=igreen, fill=igreen, legend entry=\textsc{Orig}] coordinates {
        (13.2, 0.2130871081448403) (33.599999999999994, 0.5163586351810504) (53.99999999999999, 0.20774782069701972) (74.39999999999999, 0.04417146888651587) (94.8, 0.012329900106785748) (115.19999999999999, 0.003884118143869527) (135.6, 0.001554247257547811) (156.0, 0.0005834677215804291) (176.39999999999998, 1e-06) (196.8, 0.0002922338607902146)
    };
    \addplot [mark=none, color=rred, fill=rred, legend entry=\textsc{DP-Gen}] coordinates {
        (13.2, 1e-06) (33.599999999999994, 1e-06) (53.99999999999999, 9.807795359673818e-05) (74.39999999999999, 1e-06) (94.8, 0.9996126881856131) (115.19999999999999, 0.0002922338607902146) (135.6, 1e-06) (156.0, 1e-06) (176.39999999999998, 1e-06) (196.8, 1e-06)
    };
    
    \nextgroupplot[title=\textsc{N2C2 2008}]
    \addplot [mark=none, color=bblue, fill=bblue,] coordinates {
        (331.3, 0.9548882180451128) (943.9000000000001, 0.03007618796992481) (1556.5, 0.007519796992481203) (2169.1000000000004, 1e-06) (2781.7, 0.007519796992481203) (3394.3, 1e-06) (4006.9, 1e-06) (4619.5, 1e-06) (5232.1, 1e-06) (5844.700000000001, 1e-06)
    };

    \addplot [mark=none, color=igreen, fill=igreen] coordinates {
        (331.3, 0.00967841935483871) (943.9000000000001, 0.16451712903225807) (1556.5, 0.32419454838709677) (2169.1000000000004, 0.27742035483870964) (2781.7, 0.12580745161290321) (3394.3, 0.05967841935483871) (4006.9, 0.022581645161290322) (4619.5, 0.01129132258064516) (5232.1, 0.0032268064516129034) (5844.700000000001, 0.0016139032258064515)
    };
    \addplot [mark=none, color=rred, fill=rred] coordinates {
        (377.85, 0.0032268064516129034) (985.5500000000001, 0.008065516129032257) (1593.25, 0.014517129032258064) (2200.9500000000003, 0.012904225806451612) (2808.65, 0.9612913225806452) (3416.3500000000004, 1e-06) (4024.05, 1e-06) (4631.75, 1e-06) (5239.450000000001, 1e-06) (5847.15, 1e-06)
    };
    
    \nextgroupplot[title=\textsc{PsyTAR}]
    \addplot [mark=none, color=bblue, fill=bblue] coordinates {
        (11.65, 0.03792593613676557) (30.950000000000003, 0.11397231067007271) (50.25, 0.16368738239339753) (69.55000000000001, 0.5419542324621733) (88.85, 0.01139813106700727) (108.15, 0.024760284731774416) (127.45, 0.031834366083709965) (146.75, 0.07074181351935548) (166.05, 0.003734542935743761) (185.35000000000002, 1e-06)
    };

    \addplot [mark=none, color=igreen, fill=igreen] coordinates {
        (11.65, 0.6840464723637789) (30.950000000000003, 0.2636231089768718) (50.25, 0.04135733085064681) (69.55000000000001, 0.007057056448451587) (88.85, 0.0023530188161505292) (108.15, 0.0003930031360250882) (127.45, 0.0005890047040376323) (146.75, 0.0003930031360250882) (166.05, 1e-06) (185.35000000000002, 0.0001970015680125441)
    };

    \addplot [mark=none, color=rred, fill=rred] coordinates {
        (11.65, 1e-06) (30.950000000000003, 1e-06) (50.25, 1e-06) (69.55000000000001, 1e-06) (88.85, 0.021561172481379853) (108.15, 0.9784408275186202) (127.45, 1e-06) (146.75, 1e-06) (166.05, 1e-06) (185.35000000000002, 1e-06)
    };
    \legend{\textsc{Aug-PE}, \textsc{Original}, \textsc{DP-Gen}}
    \end{groupplot}
\end{tikzpicture}
    }
    \caption{Probability density functions at of the lengths of original and synthetic datasets at $\epsilon=\infty$ as estimated by a 10-bin histogram.}
    \label{fig:text-len-distribution}
\end{figure}

\begin{figure}[!b]
    \centering
        \begin{minipage}{\textwidth}
        \centering
        \captionof{table}{Channel-wise pixel value distribution KL divergences for different approaches and privacy budgets ($\epsilon$) across image datasets.}
        \resizebox{0.70\textwidth}{!}{    

\begin{tabular}{l c c c c c}
\toprule
 & {$\epsilon = \infty$} & {$\epsilon = 4$} & {$\epsilon = 2$} & {$\epsilon = 1$} & {$\epsilon = 0.5$} \\
Method & KL-div R|G|B$\downarrow$ & KL-div R|G|B$\downarrow$ & KL-div R|G|B$\downarrow$ & KL-div R|G|B$\downarrow$ & KL-div R|G|B$\downarrow$ \\
\midrule
\rowcolor{black!10}
\multicolumn{6}{c}{\textsc{FracAtlas}} \\
Original & \scoretriple{0.00}{0.00}{0.00} & & & & \\
DPSDA & \scoretriple{1.00}{0.87}{0.97} & \scoretriple{0.74}{0.67}{0.74} & \scoretriple{1.00}{0.89}{0.99} & \scoretriple{1.03}{0.95}{1.04} & \scoretriple{0.89}{0.91}{1.00} \\
DP-MEPF & \scoretriple{0.15}{0.15}{0.15} & \scoretriple{0.42}{0.41}{0.51} & \scoretriple{0.62}{0.57}{0.85} & \scoretriple{0.50}{0.34}{0.94} & \scoretriple{0.92}{0.67}{1.44} \\
\midrule
\rowcolor{black!10}
\multicolumn{6}{c}{\textsc{SIPaKMeD}} \\
Original & \scoretriple{0.00}{0.00}{0.00} & & & & \\
DPSDA & \scoretriple{0.29}{7.05}{2.35} & \scoretriple{0.34}{6.57}{1.75} & \scoretriple{0.24}{4.38}{2.03} & \scoretriple{0.54}{7.29}{2.26} & \scoretriple{0.33}{6.52}{1.49} \\
DP-MEPF & \scoretriple{0.07}{0.07}{0.07} & \scoretriple{0.60}{0.39}{0.90} & \scoretriple{1.74}{0.79}{2.02} & \scoretriple{2.76}{1.23}{2.77} & \scoretriple{2.16}{0.96}{2.57} \\
\bottomrule
\end{tabular}}
        \label{tab:img-fidelity-2}
    \end{minipage}
    \begin{minipage}{\textwidth}
        \centering
        \resizebox{0.85\linewidth}{!}{\input{INCFIGS/figure-pixel-values}}
        \caption{Probability density functions of the per-channel pixel values at $\epsilon=\infty$ (normalised to $(0,1)$) for \textsc{FracAtlas} above and \textsc{SIPaKMeD} below.}
        \label{fig:img-pixel-distribution}
    \end{minipage}
\end{figure}

\paragraph{Fidelity:} The fidelity of the synthetic data suffers from the shift to an unknown domain. For example for text, we cannot replicate the findings of \citet{Xie2024DifferentiallyText} where the text length distributions eventually match those of the real data: The DP-GENERATOR method generates texts almost exclusively up to the generation limit. Meanwhile AUG-PE's text lengths show more variability, but they still do not match the distribution of original data (see Figure~\ref{fig:text-len-distribution}). 

Similarly, the measured similarity between original and synthetic text is strikingly low, as reported in Table~\ref{tab:text-fidelity} and exemplified in Figure~\ref{fig:text-example} in the Appendix. While MAUVE scores are close to 1 between original train and test splits, they're close to 0 between synthetic and real data. For context, the MAUVE score between two unrelated (real) datasets, \textsc{PsyTAR} and \textsc{Hallmarks Of Cancer} is $0.004$, only marginally lower than the scores reported between real and synthetic data on the same dataset. Encouragingly, the difference is less noticeable (albeit present) when looking at named entities as a proxy for semantic content and n-gram frequency (reported in Table~\ref{tab:text-fidelity-2}) as a measure for lexical diversity, as measured by KL divergence of the respective distributions in real and synthetic data, suggesting that the differences are rooted in discourse, coherence, style or syntax rather than being vocabulary-based.

For images we observe similar trends: the FID scores, reported in Table~\ref{tab:img-fidelity} are much higher than those reported in literature on popular datasets like CIFAR~\cite{Lin2024DifferentiallyImages,Harder2023Pre-trainedGeneration}.

Likewise, channel-wise pixel value distributions differ between original and synthetic data, even without privacy guarantees, as shown in Table~\ref{tab:img-fidelity-2} and Figure~\ref{fig:img-pixel-distribution}. Here, data generated by DP-MEPF largely matches the original distribution, without incorporating more intricate spikes in pixel values (as is the case with e.g., \textsc{FracAtlas}), which makes sense, as without privacy guarantees the training regime is similar to that of a GAN, which are known to do well on image generation tasks. Looking at qualitative examples of the \textsc{FracAtlas} dataset presented in Figure~\ref{fig:img-example}, it becomes apparent that for DP-MEPF the image quality deteriorates significantly with added noise. Meanwhile, DPSDA's image quality is strikingly good, but the underlying generative model struggles with generating realistic x-ray images, resulting in images that often do not resemble any human body part at all.

\begin{wrapfigure}{R}{0.45\textwidth}
    \vspace{-1.5em}
    \centering
    \captionof{table}{FID scores for different approaches and privacy budgets ($\epsilon$) across image datasets.}
    \label{tab:img-fidelity}
    \resizebox{0.45\textwidth}{!}{

\begin{tabular}{lrrrrr}
\toprule
Method & {$\epsilon = \infty$} & {$\epsilon = 4$} & {$\epsilon = 2$} & {$\epsilon = 1$} & {$\epsilon = 0.5$} \\
 & {FID $\downarrow$} & {FID $\downarrow$} & {FID $\downarrow$} & {FID $\downarrow$} & {FID $\downarrow$} \\
\midrule
\rowcolor{black!10}
\multicolumn{6}{c}{\textsc{FracAtlas}} \\
Original & $21.31$ &  &  &  &  \\
DPSDA & $139.02$ & $138.34$ & $149.22$ & $157.61$ & $190.03$ \\
DP-MEPF & $52.37$ & $206.70$ & $223.88$ & $342.40$ & $406.98$ \\
\midrule
\rowcolor{black!10}
\multicolumn{6}{c}{\textsc{SIPaKMeD}} \\
Original & $31.84$ &  &  &  &  \\
DPSDA & $288.46$ & $264.41$ & $285.56$ & $254.03$ & $259.25$ \\
DP-MEPF & $52.86$ & $365.88$ & $427.30$ & $479.78$ & $489.80$ \\
\bottomrule
\end{tabular}}
    \vspace{-1.5em}
\end{wrapfigure}
\sloppy

\paragraph{Summary of Findings}
Overall, our findings present credible empirical evidence for the challenges outlined in Section~\ref{sec:challenges}. We show that synthetic data generation approaches relying on foundational models pre-trained on large-scale datasets indeed struggle to transfer the performance reported on open-domain datasets to specialised domains, such as biomedical texts and images. Indeed, the issues seem to stem from the lack of underlying models' exposure to specialist domains rather than the privatisation mechanisms per say, as the quality of generated data deteriorates less with stricter privacy bounds compared to the deterioration attributed to the domain shift. 

Our findings motivate the need for a benchmark to evaluate existing approaches for synthetic data generation in a more realistic setting---i.e., dealing with complex data from domains unseen during foundational model pre-training. We further show that while the overall results as measured by popular fidelity metrics and down-stream performance are largely negative, different aspects are impacted differently, e.g., the overlap between recognised entities in texts is relatively high. This calls for a multi-faceted evaluation to understand the limitations of the approaches in detail, thus allowing for targeted improvements.

\section{Conclusion}
\label{sec:conclusion}
Privacy-preserving synthetic data generation represents a promising approach to bridge the gap between the need for data sharing and strict privacy requirements in high-stakes domains. Our survey has examined the theoretical foundations, methodologies, and evaluation frameworks across different data modalities, highlighting the significant progress made in recent years through the integration of generative AI with differential privacy guarantees. However, our empirical analysis reveals that current state-of-the-art methods face substantial challenges when applied to specialized domains like healthcare, with performance degrading significantly under realistic privacy constraints ($\epsilon \leq 4$). This performance gap underscores the limitations of current approaches when transferring from general to specialized domains, particularly when utilizing foundation models pre-trained on web-scale data. We thus provide concrete evidence for hypotheses presented in literature~\cite{Tramer2024Position:Pretraining}.

Looking ahead, we identify several critical research directions that could advance the field: developing benchmarks that realistically represent high-stakes domains without compromising privacy; creating robust evaluation frameworks that effectively measure both utility and privacy leakage; addressing multimodal data generation to reflect real-world data complexity; and strengthening empirical privacy verification through more rigorous membership inference attacks. These advancements will be essential for synthetic data to fulfill its potential in enabling broader access to sensitive data while maintaining robust privacy guarantees. 

\section{Acknowledgement}
This research is part of the IN-CYPHER programme and is supported by the National Research Foundation, Prime Minister’s Office, Singapore under its Campus for Research Excellence and Technological Enterprise (CREATE) programme. We are grateful for the support provided by Research IT in form of access to the Computational Shared Facility at The University of Manchester and the computational facilities at the Imperial College Research Computing Service\footnote{DOI: \url{https://doi.org/10.14469/hpc/2232}}.

\bibliography{references, references2}

\bibliographystyle{unsrtnat}
\appendix
\section{More details for the KDE example}

Let \( \{x_1, x_2, \ldots, x_N\} \) be vector samples from a distribution with an unknown density \( f(x) \). The kernel density estimator is given by:
\[
\hat{f}(x|h) = \frac{1}{N \cdot h} \sum_{n=1}^N K\left(\frac{x - x_n}{h}\right),
\]
where \( K(\cdot) \) is the kernel function and \( h > 0 \) is the bandwidth parameter.

We wish to understand the maximum change in the kernel density estimator when we remove a point from the dataset. Removing a point from the dataset is perhaps the most obvious way we have of directly indicating whether the presence of an individual in the dataset can be equipped with a clear privacy bound; if we cannot tell whether a single specific known item is in the dataset or not, then we can safely say that a membership inference attack will likely fail. 

\subsection{$N=1$}
We frame this treatment around the two-dimensional toy embedding space example, presented in Section~\ref{sec:background}. If we have only one individual in the dataset, with embedding vector $x_1 \in \mathbb{R}^2$, then the kernel density estimate is simply the kernel function ``centered'' at that point in space:
\[
\hat{f}(x|h) = \frac{1}{h} \sum_{n=1}^N K\left(\frac{x - x_n}{h}\right),
\]
For a Gaussian kernel, we take the $K$ to be the standard normal distribution; we use the Euclidean length of vector difference $x - x_n$  in embedding space to determine the kernel mapping. The bandwidth parameter, $c=\sigma_k^2$ is, then corresponds to the variance of the general Gaussian kernel with mean 0 and spread $\sigma_k$. The peak height of the kernel is therefore $1/{2\pi \sigma_k^2}$ = $1/(2\pi c)$.

\subsection{The case for $N=2$}
We can deduce the maximum change in the kernel density estimate to removing a point by considering a simple case of two individuals in the embedding space example introduced in the main body of the paper. We have two individuals in dataset $D$, represented by the two-dimensional embedding vectors \( x_1 \) and \( x_2 \), and we wish to understand the maximum change in the kernel density estimate when we remove \( x_2 \) from the dataset to create $D^{\prime}$. We can write the kernel density estimate as:
\[
\hat{f}(x|h) = \frac{1}{2 \cdot h} \left[ K\left(\frac{x - x_1}{h}\right) + K\left(\frac{x - x_2}{h}\right) \right].    
\]
We can then write the difference between two kernel density estimates, one with both points and the other with only one point as:
\begin{eqnarray}
\max_{x,x_1,x_2} \left| \hat{f}(x|h) - \hat{f}(x|h, x_2) \right| &=& \max_{x,x_1,x_2} \left| \frac{1}{2 \cdot h} \left[ K\left(\frac{x - x_1}{h}\right) + K\left(\frac{x - x_2}{h}\right) \right] - \frac{1}{h} K\left(\frac{x - x_1}{h}\right) \right| \nonumber \\
&=& \frac{1}{2 \cdot h} \max_{x,x_1,x_2} \left | K \left(\frac{x - x_2}{h} \right ) - K\left(\frac{x - x_1}{h}\right) \right | 
\label{eq:app1}
\end{eqnarray}

Equation~(\ref{eq:app1}) expresses the idea that we must find the overall maximum as we vary $x$ in evaluating the kernel density, and do so for all possible embedded two-sample datasets $x_1$ and $x_2$. The maximum absolute differences will occur when the two shifted kernel functions -- effectively separated by the Euclidean distance between them -- will occur as $d_E(x_1, x_2) \rightarrow \infty$. As a function of $x$, these maxima will also occur at the points $x=x_1$ and $x=x_2$ (that is, at the value of $x$ corresponding to the two embedding points, themselves). 

The values of the {\em differences}, i.e. ignoring the absolute values, allows us to apply simple partial derivatives to determine conditions of maxima and minima in $x_1$ and $x_2$. These will correspond to $\pm$ half the height of each of the Gaussian kernels, which will have local maxima/minima at $x = x_1$ and $x = x_2$, respectively. This is because (for example), @$x = x_1$ the height of the KDE will be slightly greater than $K(0)/2$ when both points are present, and $K(0)$ when only one point is present, giving a difference of just greater than $-K(0)/2$. On the other hand, at location $x=x_2$, the height of the KDE on $D$ will be slightly greater than $K(0)/2$, when both points are present, and just over $0$ when only one point is present $D^{\prime}$, giving a difference of just less than $K(0)/2$.  

We revert to taking the absolute value of this change in the treatment that follows. We can then bound the maximum change in the kernel density estimate as:
\begin{eqnarray}
|\delta_f| < \frac{1}{4 \pi \sigma_k^2}
\end{eqnarray}
where $\sigma_k$ is the standard deviation of the kernel. Similar arguments follow if remove  $x_1$ instead of $x_2$.

\subsection{The case for general $N$}
The case for $N$ follows on with a general expression of:
\begin{eqnarray}
    |\delta_f| < \frac{1}{2 \pi N \sigma_k^2}
\end{eqnarray}
These results are used to determine the bounds for the illustration of Figure~\ref{fig:DPProf} in the main body of the paper, using $\sigma_k = 0.2$, and $N=26$. Note that this choice of $\sigma_k$ is below what one would typically use to provide a smooth kernel density estimate, but it has been chosen to clearly illustrate the effectiveness of the privacy preserving mechanisms on the distributions. Indeed, using a larger value of $\sigma_k$ provides such a smoothing effect over the embedding space that the nature of the change is quite small if we remove a point. Thus, one's {\em practical} mileage in applying DP for privacy preservation may vary depending on the bandwidth for the KDE, and on the sparsity of points in the (embedding) neighbourhood of record removal.

Tighter bounds can be supplied if the embedding space is bounded; for example, it might be that an encoder network uses a sigmoid activation function, which bounds the embedding space to $[0,1]$. In this case, because the maximum distance between two embedding vectors is bounded by $\sqrt{2}$, the maximum change in the kernel density estimate will also be bounded; the result of density estimation is then non-trivial, because it will depend on whether the KDE output is also bounded and renormalised accordingly.

\subsection{A note on the choice of KDE parameters}
To demonstrate the effect of the Laplacian mechanism on kernel density estimation, we used a couple of specific Gaussian kernel parameters for the KDE plots shown in the main paper. These were selected to best illustrate the effects of the privacy preserving mechanisms on the distributions, and also the Catch-22 situation of loss of detail in distribution. For Figure~\ref{fig:KDE}, we used $\sigma_k=0.4$, illustrating a density estimate that typifies the usefulness of kernel density estimation.  We used ($\sigma_k=0.2$) to emphasise the effect of removing a data sample, and for the bounds analysis. Although the choices were arbitrary, they are consistently applied: we used the same kernel for pairs of datasets $D$ and $D^{\prime}$ when constructing estimates of the ratio of probabilities, and their bounds. We do not -- in this situation or elsewhere -- use automatic kernel selection methods, as theseare practically useless if one wishes to compare two or more distributions estimated from (different) real-world data.

\section{Implementation Details}
We follow the original implementations and adapt them mostly at the data level. Where additional prompting is required, i.e. AUG-PE~\cite{Xie2024DifferentiallyText} and DPSDA~\cite{Lin2024DifferentiallyImages}, we follow the original methodology to adapt the prompts to our datasets accordingly. More details can be seen in the code at~\url{https://github.com/ImperialGlobalSingapore/synth-data}.

\section{Text example}
Figure~\ref{fig:text-example} shows examples of generated texts at different $\epsilon$ levels.
\begin{figure}[htbp]
    \resizebox{1\textwidth}{!}{
\begin{tabular}{|lp{15cm}|}
\hline
$\mathbb{\epsilon}$  & \textbf{Text} \\
\hline
\rowcolor{black!10}\multicolumn{2}{|c|}{Original} \\
$-$& \emph{Over a 10 yr period I have gained 70 lbs Other than that, it helped me with my anxiety attacks and depression.}\\
\hline
\rowcolor{black!10}\multicolumn{2}{|c|}{AUG-PE} \\
$\infty$& \emph{Tireless life support systems crumble at my disposal as a high stakes gamble with my vitality wears thin every time, as concerns multiply into a formidable puzzle that wears me down with each wearisome day, until that miraculous pill started to work wonders on an escalating tide of anguish, easing the turbulent depths of anxiety}\\
$4$& \emph{The initial relief from the medication has subsided for the last six months, and I started experiencing unbearable discomfort and lack of rest just one month into treatment. Or alternatively: Treatment started showing its effect just a month after that, which left me feeling irritable and struggling to get enough rest after just six short}\\
$2$& \emph{The escalating inflammation in my gastrointestinal area caused persistent and worsening symptoms, rendering my previous choice of Zetricity's treatment ineffective. or As time passed, severe gastrointestinal symptoms worsen despite my initial treatment choice Zetrics, ultimately making its impact less significant. or My previous medication Zetics, chosen to}\\
$1$& \emph{Anxious individuals often require treatment with caution and careful consideration, as medications may have varying degrees of efficacy, especially for certain conditions. However, personal experiences can diverge significantly, demonstrating that not all medications can lead to the desired results, even those aimed to address related symptoms. Various encounters can evoke emotional responses that may}\\
$0.5$& \emph{"Sorry, I'm not familiar with the issue you're referring to, so I'd be happy to help, but I need more context or specific details for me to provide a more informed solution."}\\
\hline
\rowcolor{black!10}\multicolumn{2}{|c|}{DP-GENERATOR} \\
$\infty$& \emph{Totacking with many other people comments.  Burns bouts of suicide. omicidalnt put away.  headaches changes in line with omicinations.  rage didn't seem to lessen.  migraines.  extreme sleeplessness.  racing thoughts.  shaky. nervous. decreasedensitivity. stomachache. difficulty to concentrate. lasting ones: obsessive obsess over medication.  profuse sweating and dry mouth. tantrums. decreased emotion range. lack of energy. etc Weight gain. bad}\\
$4$& \emph{No desire to control myself (Brobrewation before this meds, started to get on crowd). Liver extended, initially seemed okay, but woke up day after attempt with whole weekend bender, and with a half dose of nice red alcohol (Weeks before Brokenthrothing, back to Happy Drobie). Straight away after the dose was a complete failure, the alcohol is messing up his thinkin and myin as he was subject to ever tiny chore at the dispesn). Wanting to continue}\\
$2$& \emph{During the first 48 hours of my life on this drug, I had no energy, joint aches and body aches, sleepiness, and nothing on a good appetite. Not a worry of drug effects.
Was not the sleep I hoped for. Burning twice the muscle weight night. Viglaintness flush. Walter, I'm sleeping well and whatnot. Folds cheap and budget/riskaty fever was unnoticeable and kept. Unused controlled. Minimum of emotion, deathizhed}\\
$1$& \emph{I could not remember those days in an excited manner unlike ever. Weight gain is the only ill effect(I am not used to madness, so this weirdness around dietism). That aside, on July 9th treadmill session with workouts - 2 aborted days. Went getting and walking uphill. Diarrhoeia really. But that was ok because I am used to it. Went out and back, checked all calorie sites. Headed towards sugar free sweet depressant diet. Can't find anything}\\
$0.5$& \emph{Starting with Crestor and taking 200mg a day of Sunday's were great. One morning mind is not with me, so I had to take plus another of my two with 300mg starting on Mondays and 500mg from thnot to May. Poster il mio muta i come verificando tutti il bilanciamento intramondanei, ebbientesse inendingur deces a vuoi decenisti. Aggiugno ho dato 186mgz}\\
\hline
\end{tabular}

}
    \caption{Random generations for the label ``Adverse Drug Reaction'' at different levels of privacy guarantee $\epsilon$. Note the mismatch in text lengths and how for with increasing $\epsilon$, the quality of generations deteriorates. AUG-PE fails to filter out irrelevant answers (i.e. the model's refusal to comply with the prompt), while the DP-fine-tuned generator invents new words (``Brokenthrothing'') or suddenly changes language.}
    \label{fig:text-example}
\end{figure}
\end{document}